\documentclass[aps,prb,amsmath,amssymb,twocolumn,superscriptaddress,floatfix]{revtex4-1}
\usepackage{graphicx}
\usepackage{helvet}
\usepackage{hyperref}

\begin{document}
\title{Presence of exotic electronic surface states in LaBi and LaSb}

\author{X. H. Niu}
\author{D. F. Xu}
\author{Y. H. Bai}
\affiliation{State Key Laboratory of Surface Physics, Department of Physics,  and Advanced Materials Laboratory, Fudan University, Shanghai 200433, People's Republic of China}
\affiliation{Collaborative Innovation Center of Advanced Microstructures, Nanjing 210093, People's Republic of China}
\author{Q. Song}
\affiliation{State Key Laboratory of Surface Physics, Department of Physics,  and Advanced Materials Laboratory, Fudan University, Shanghai 200433, People's Republic of China}
\affiliation{Collaborative Innovation Center of Advanced Microstructures, Nanjing 210093, People's Republic of China}
\author{X. P. Shen}
\author{B. P. Xie}
\affiliation{State Key Laboratory of Surface Physics, Department of Physics,  and Advanced Materials Laboratory, Fudan University, Shanghai 200433, People's Republic of China}
\affiliation{Collaborative Innovation Center of Advanced Microstructures, Nanjing 210093, People's Republic of China}
\author{Z. Sun}
\affiliation{Collaborative Innovation Center of Advanced Microstructures, Nanjing 210093, People's Republic of China}
\affiliation{National Synchrotron Radiation Laboratory, University of Science and Technology of China, Hefei, Anhui 230029, People's Republic of China}
\author{Y. B. Huang}
\affiliation{Shanghai Synchrotron Radiation Facility, Shanghai 201204, People's Republic of China}
\author{D. C.\ Peets}\email{dpeets@fudan.edu.cn}
\author{D. L. Feng}\email{dlfeng@fudan.edu.cn}
\affiliation{State Key Laboratory of Surface Physics, Department of Physics,  and Advanced Materials Laboratory, Fudan University, Shanghai 200433, People's Republic of China}
\affiliation{Collaborative Innovation Center of Advanced Microstructures, Nanjing 210093, People's Republic of China}

\date{\today}

\begin{abstract}
  Extremely high magnetoresistance (XMR) in the lanthanum monopnictides La$X$ ($X$ = Sb, Bi) has recently attracted interest in these compounds as candidate topological materials.  However, their perfect electron-hole compensation provides an alternative explanation, so the possible role of topological surface states requires verification through direct observation.  Our angle-resolved photoemission spectroscopy (ARPES) data reveal multiple Dirac-like surface states near the Fermi level in both materials. Intriguingly, we have observed circular dichroism in both surface and near-surface bulk bands. Thus the spin-orbit coupling-induced orbital and spin angular momentum textures may provide a mechanism to forbid backscattering in zero field, suggesting that surface and near-surface bulk bands may contribute strongly to XMR in La$X$. The extremely simple rock salt structure of these materials and the ease with which high-quality crystals can be prepared suggests that they may be an ideal platform for further investigation of topological matter.

\end{abstract}



\maketitle

\section{Introduction}

The discovery of topologically nontrivial quantum states in condensed matter systems, such as in topological insulators\cite{bernevig2006TI, zhang2009TI, xia2009TI, fu2007TI, fu2007TI2, Teo2008TI, hsieh2008TI, chen2009BiTe}, Dirac semimetals\cite{Young2012dirac,Liu2014Na3Bi, Liu2014Cd3As2, Xu2015dirac}  and Weyl semimetals\cite{Wan2011weyl, Burkov2011weyl, Weng2015weyl, Lu2015weyl, Xu2015TaAs, Lv2015TaAs, Yang2015weyl, Huang2015weyl, Xu2015weyl}, provides a platform for investigating particles with Dirac-like linear dispersions.  This allows testing of particle physics predictions in crystalline solids, while the topologically protected nature of the states may prove useful for spintronics.  Many of these materials also exhibit extremely high magnetoresistance (XMR), with potential applications in reading out magnetically-stored data.
Recently, the lanthanum monopnictides La$X$ ($X$ = P, As, Sb and Bi) were predicted to be topological insulators\cite{Fuliang2015}. This inspired an explosion of mainly transport work on LaSb\cite{Kasuya1993,Tafti2016,Zeng2016}, LaBi\cite{Kasuya1993,Stepanov2015,Sun2016,Wu2016,Kumar2016}, and YSb\cite{Ghimire2016,Yu2016,Pavlosiuk2016}, all of which crystallize in the well-known NaCl structure. All show a large unsaturated magnetoresistance, which is often attributed to the quantum limit of the Dirac fermions \cite{Abrikosov1998}, but may also arise from complicated factors such as electron-hole compensation \cite{Yang1999, Mun2012}. A recent angle-resolved photoemission spectroscopy (ARPES) study on WTe$_2$ suggested that its anomalously large magnetoresistance may be attributable to spin and orbital angular momentum textures which would suppress backscattering of the quasiparticles\cite{Jiang2015}. Similar to the case of Cd$_3$As$_2$\cite{liang2015CdAs}, such a mechanism would be progressively invalidated under a magnetic field, thus causing large magnetoresistance. 
First-principles calculations of LaBi and LaSb based on the two-band model, however, claimed that these materials are topologically trivial and their extremely high magnetoresistance could be attributed to perfect electron-hole compensation, without recourse to topologically nontrivial states\cite{Guo2016}.

Recent ARPES experiments indicate that LaSb is topologically trivial and its properties are well explained by electron-hole compensation\cite{Zeng2016}. On the other hand, ARPES experiments on LaBi show linear band dispersion\cite{Wu2016} and indicate that LaBi may host an odd number of surface Dirac cones\cite{Nayak2016}, making it topologically nontrivial. One Dirac cone is located at the surface Brillouin zone (SBZ) center while the other two are found at the SBZ corner\cite{Nayak2016}. In YSb, meanwhile, the negative Hall coefficient indicates that electrons are the dominant carriers, calling into question the perfect compensation of electrons and holes\cite{Yu2016}. 

Unfortunately, the Dirac cones in LaBi were not very clearly resolved\cite{Nayak2016}, necessitating a closer investigation of its band structure and suggesting that a re-examination of LaSb would also be desirable. In this paper, we report the surface and bulk electronic structures of LaBi and LaSb by ARPES. We identify a clear band anti-crossing along the $\bar{\varGamma}$--$\bar{X}$ direction of LaBi, which perfectly matches with the calculated results\cite{Nayak2016}. We show clear evidence that LaBi hosts one Dirac cone at the SBZ center and two Dirac-cone-like surface bands at the zone corner. The results on LaSb differ with the previous ARPES data\cite{Zeng2016} --- we do find evidence of Dirac-cone-like surface bands. Moreover, both surface and near-surface bulk bands exhibit circular dichroism (CD). Our results unveil the exotic surface states in LaBi and LaSb, which might be topologically nontrivial in nature. In addition to electron-hole compensation, forbidden backscattering in surface and near-surface bulk bands likely contributes to the low zero-field resistivity, which would give rise to anomalous XMR when applying magnetic field in LaBi and LaSb.

\begin{figure*}[htb] 
\includegraphics[width=\textwidth]{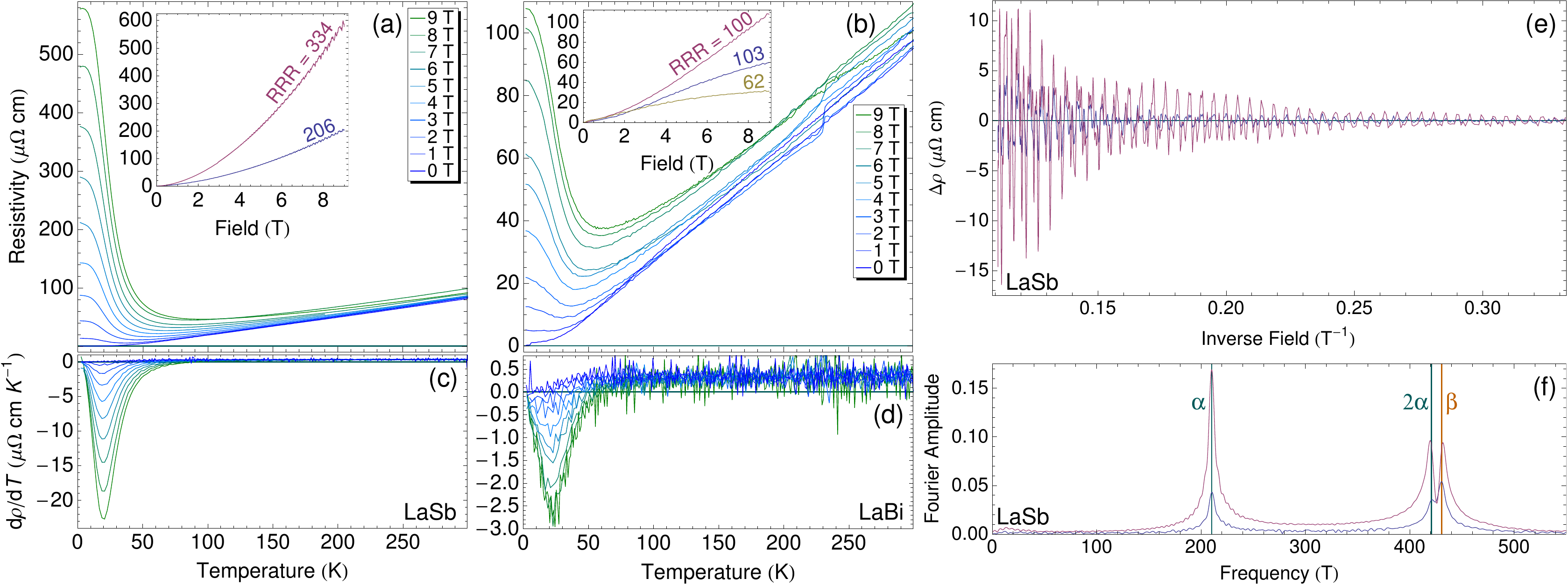}
\caption{\label{transport}Transport results on LaSb and LaBi.  Resistivity of (a) LaSb and (b) LaBi under various fields parallel to (001).  These samples had residual resistivity ratios (RRRs) of 334 and 100, respectively.  The insets compare the field-dependence of selected samples at 1.8\,K, and demonstrate clear Shubnikov-de~Haas oscillations.  (c) and (d) show the temperature derivatives of the data in panels a and b, respectively.  (e) Quantum oscillations were clearly visible in LaSb down to low fields at 1.8\,K, and implied at least two distinct frequencies as shown in panel f [colors as for the inset to panel a].}
\end{figure*}

\section{Experimental}

Black, blocky crystals of typical dimension $5\times 5\times 5$\,mm$^3$ were grown from metal fluxes, closely following established techniques for this family of materials\cite{Canfield1991}. To grow LaSb, La metal powder (Alfa Aesar, 99.9\%), Sb spheres (Alfa Aesar, 99.999\%), and Sn pieces (Aladdin, 99.999\%) were sealed under vacuum in a quartz tube, with atomic ratio 1.5:1:20.  The small La excess was used to compensate for loss by reaction to the quartz.  To grow LaBi, La powder, Bi powder (Alfa, 99.5\%), and In pieces (Alfa, 99.99\%) were sealed under vacuum in a quartz tube, with atomic ratio 1:1:20.  The mixtures were heated to 1050$^\circ$C, held for 2 hours, cooled over the course of 5--7~days to 750$^\circ$C, then cooled freely to room temperature.  A small temperature gradient was applied, with the bottom end of the quartz tube cooler, to encourage crystallization.  The resulting solidified ingot was then remelted at 350$^\circ$C and the tin or indium flux was centrifuged off, revealing the crystals.  In both cases, the first crystals obtained were large and of excellent quality, so no further attempts to optimize the growth process were made.  

Resistivity measurements were performed between 1.8 and 300\,K in
fields up to 9\,T in a Quantum Design PPMS by a standard four-probe
technique, with a drive current of 8\,mA; for each compound, several
pieces of the same crystal were measured.  Quantum
oscillations were isolated by subtracting a quartic polynomial fit
from the field-dependent resistivity data, before performing a Fourier
transform. Since both Sn and In superconduct within the 
measurement window, the resistivity also provides a test for remnant 
metal flux.  

High-resolution ARPES measurements were performed at beamline 5-4 of the Stanford Synchrotron Radiation Lightsource (SSRL), using Scienta R4000 electron analyzers.  The experimental geometry is depicted in Fig.~\ref{FS}(c). The overall energy resolution was 10\,meV, and the angular resolution was 0.3$^\circ$. All samples were cleaved \textit{in situ} under ultrahigh vacuum conditions and measured at temperatures less than 15\,K. During measurements, the spectroscopy qualities were carefully monitored to avoid degradation of the data due to sample aging. 

\section{Results}

\begin{figure*}[htb] 
\includegraphics[width=\textwidth]{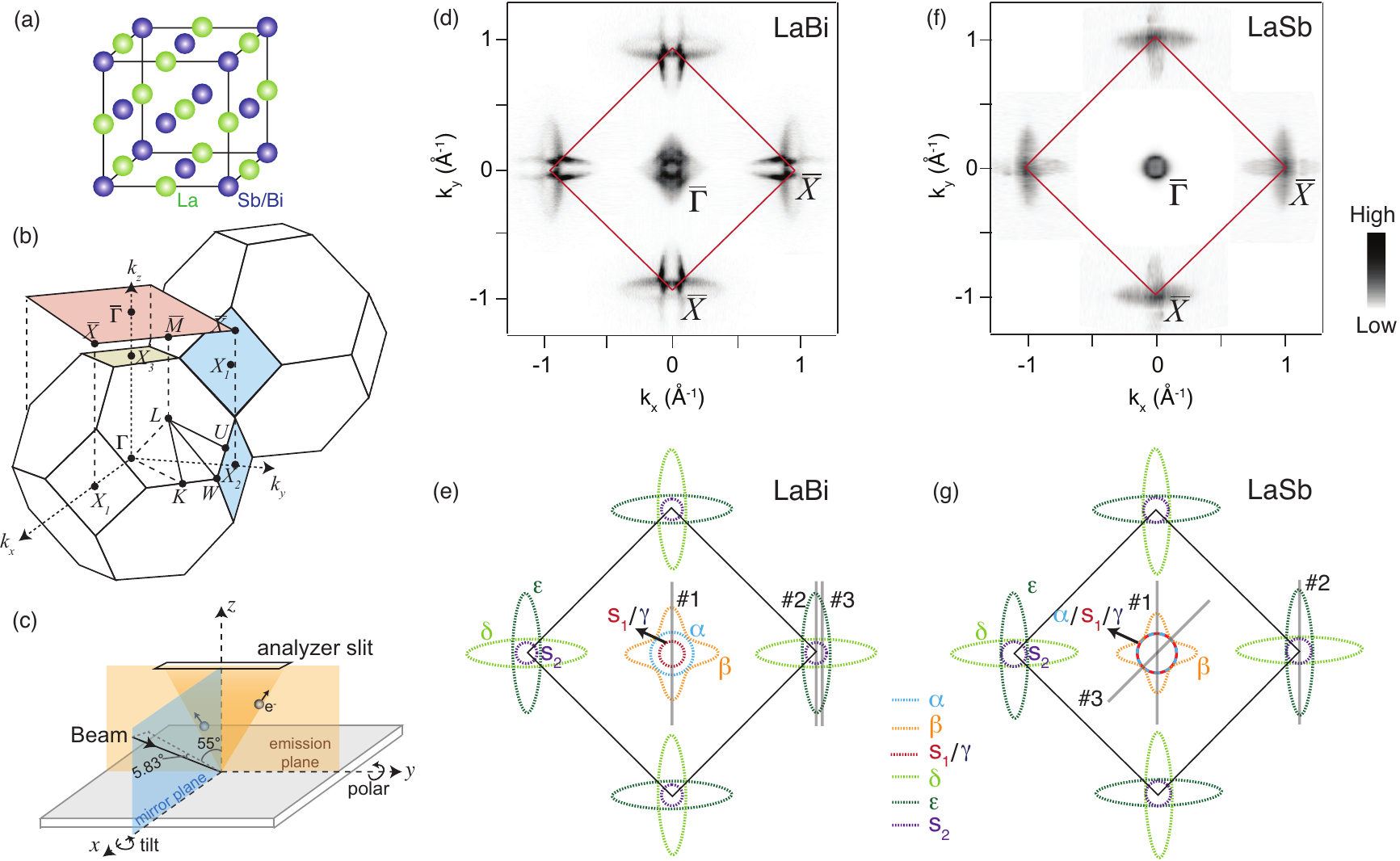}
\caption{\label{FS} (a) Crystal structure of LaBi and LaSb. (b) Brillouin zone of bulk LaSb/LaBi and the (001)-projected surface Brillouin zone. (c) The ARPES experimental setup. The analyzer slit is vertical to the mirror plane. The emission plane is defined by the analyzer slit and the sample surface normal. The angle between the direction of the beam and the emission plane is $5.83^{\circ}$. In the emission plane, the angle between the projection of the beam direction and the sample normal direction is $55^{\circ}$. (d), (f) False-color plots of the photoemission intensity at the Fermi energy ($E_F$) of LaBi and LaSb, respectively, integrated over the energy window ($E_F-15$\,meV, $E_F + 15$\,meV). These Fermi surface maps have been fourfold-symmetrized. The intensity around $\bar{X}$ of LaSb has been enhanced to compensate for the weak signal.  (e), (g) Schematic representations of the Fermi surfaces in panels d and f, respectively; pocket sizes have been exaggerated for clarity. Data in panels d and f were taken with 30 eV linearly-polarized and 24 eV circularly-polarized photons, respectively.}
\end{figure*}

\begin{figure*}[htb] 
\includegraphics[width=\textwidth]{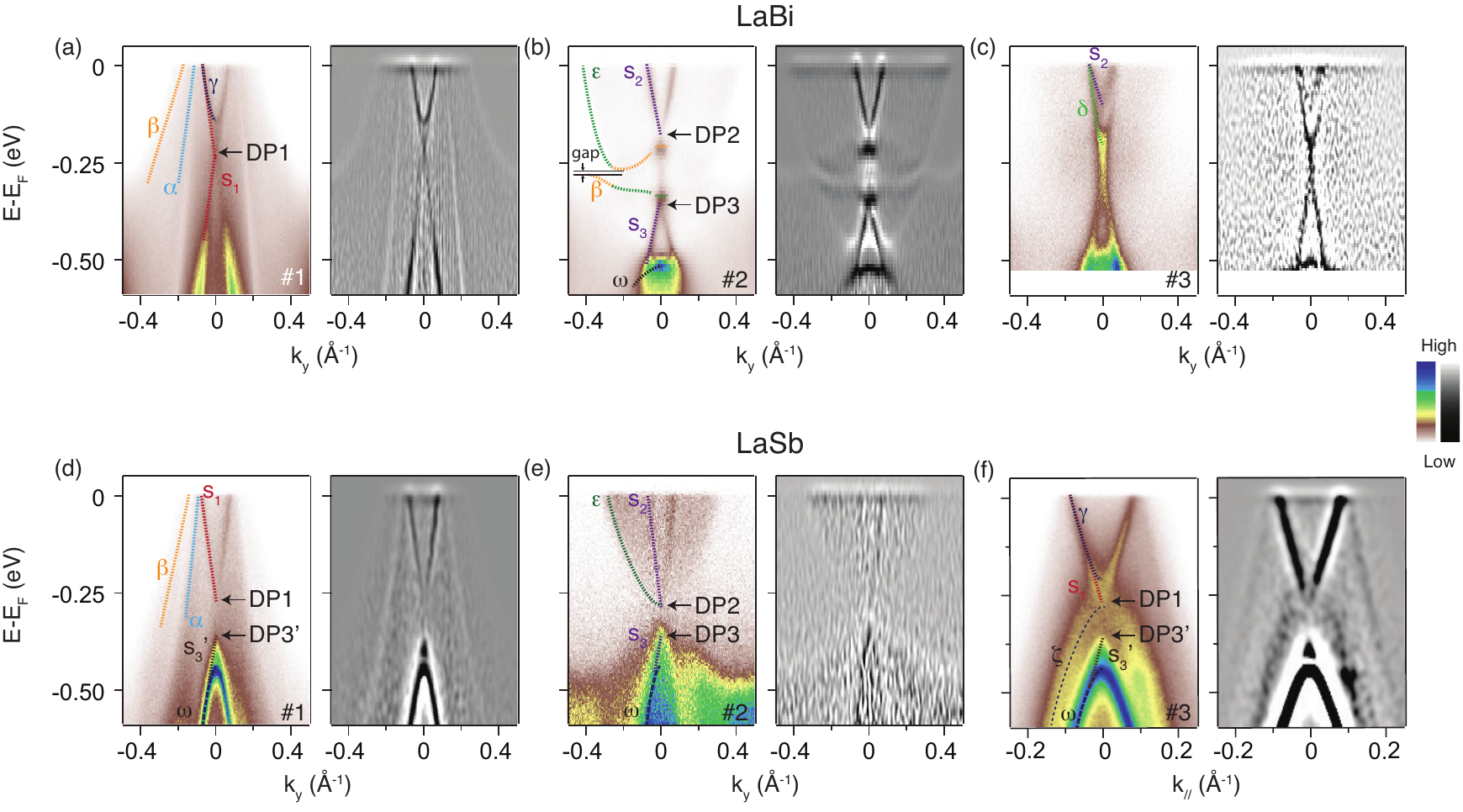}
\caption{\label{bands}Band structures of LaBi and LaSb. (a-c) Photoemission intensity $I(k, \omega)$ of LaBi and its corresponding 2D curvature intensity plot along the $\bar{\varGamma}$--$\bar{X}$ direction [cuts \#1 -- 3, respectively, in Fig.~\ref{FS}(e)]. (d-f) Photoemission intensity $I(k, \omega)$ of LaSb and its corresponding 2D curvature intensity plot along cuts \#1 -- 3, respectively, in Fig.~\ref{FS}(g)]. The Greek letters indicate  bulk bands, while $s_1$ through $s_3$ label surface bands. Data in panels a-c and d-f were taken with 22 eV and 26 eV photons, respectively.}
\end{figure*}

The resistivity $\rho$ of both LaSb and LaBi, shown in Fig.~\ref{transport}, was measured to verify that our samples behaved as in other reports.  The residual resistivity ratio RRR = $\rho$(300\,K)/$\rho(T\rightarrow 0$\,K), which often serves as a measure of sample quality, varied significantly between different pieces of the same crystal, but in all cases was high.  At 1.8\,K, Shubnikov-de~Haas oscillations were clearly visible at least as low as 3\,T, indicative of the remarkably high crystal quality readily obtained in these materials.  The highest-RRR crystal of each material showed a possible drop in resistivity at low temperatures, suggestive of remnant flux; however, the quantum oscillations were strongest in these samples, indicating that the RRR reflects the intinsic sample quality.  The suspect low-temperature points were neglected in all analysis.  The lowest quantum oscillation frequencies in LaSb of 210.4 and 430.8\,T correspond to extremal areas perpendicular to the cubic (001) axis of 2.008 and 4.112\,nm$^2$, respectively.  Quantum oscillations were not investigated in detail, since comprehensive angle-dependent studies are already available on both LaBi\cite{Yoshida2001,Kitazawa1983,Hasegawa1985} and LaSb\cite{Kitazawa1983,Hasegawa1985,Settai1993,Yoshida2000} and our frequencies are fully consistent with previous reports.  Higher magnetoresistance was generally obtained for samples with smaller cross-sections, as would be expected if the surface and bulk provided separate transport channels, but the number of samples measured does not allow for a reliable conclusion in this regard.  In such a scenario, the contribution from topologically-protected surface states would make the residual resistivity ratio less useful for characterizing sample quality.  As can be seen in Figs.~\ref{transport}(c) and (d), the highest slope in the low-temperature upturn remains at constant temperature.  The transport results presented here are consistent with those of other groups\cite{Kitazawa1983,Hasegawa1985,Settai1993,Kasuya1993,Stepanov2015,Tafti2016,Zeng2016,Sun2016,Wu2016,Kumar2016}, and serve to demonstrate that the crystals on which we report ARPES behave exactly as expected.  

Having verified that our crystals exhibit the same magnetoresistance upturn that recently piqued interest in these materials, we now turn to ARPES. We note that ARPES probes the electronic structure near the surface, where the atoms have different coordination than in the bulk, and that electron density in surface states must come at the expense of bulk bands --- we use ``bulk bands" to refer to the near-surface bulk-like bands.


At ambient pressure, LaBi and LaSb crystallize in the simple rock salt structure as illustrated in Fig.~\ref{FS}(a). The corresponding Brillouin zone (BZ) and the (001)-projected SBZ are depicted in Fig.~\ref{FS}(b). The Fermi surfaces of LaBi and LaSb in the $k_x$-$k_y$ plane are shown in Figs.~\ref{FS}(c) and (e) respectively, and their corresponding schematic representations are depicted in panels (d) and (f), respectively. The Fermi surface topologies are basically consistent with calculations\cite{Hasegawa1985,Guo2016}, including one electron pocket at the BZ corner (the $\delta$ band) and two hole pockets at the BZ center (the $\alpha$ and $\beta$ bands). The perpendicular $\delta$ and $\varepsilon$ pockets in Figs.~\ref{FS}(c-f) are from two inequivalent $X$ points ($X_2$ and $X_1$, in blue), and appear together due to the poor $k_z$ resolution of our experiments using vacuum ultraviolet photons (VUV-ARPES) --- the bands from a specific $k_z$ plane may have a projection over a wide range of $k_z$\cite{Song2016,Xu2016}. Meanwhile, the $X_3$ point (beige) is projected to the $\bar{\varGamma}$ point for the same reason. 

\begin{figure*}[htb]
\includegraphics[width=\textwidth]{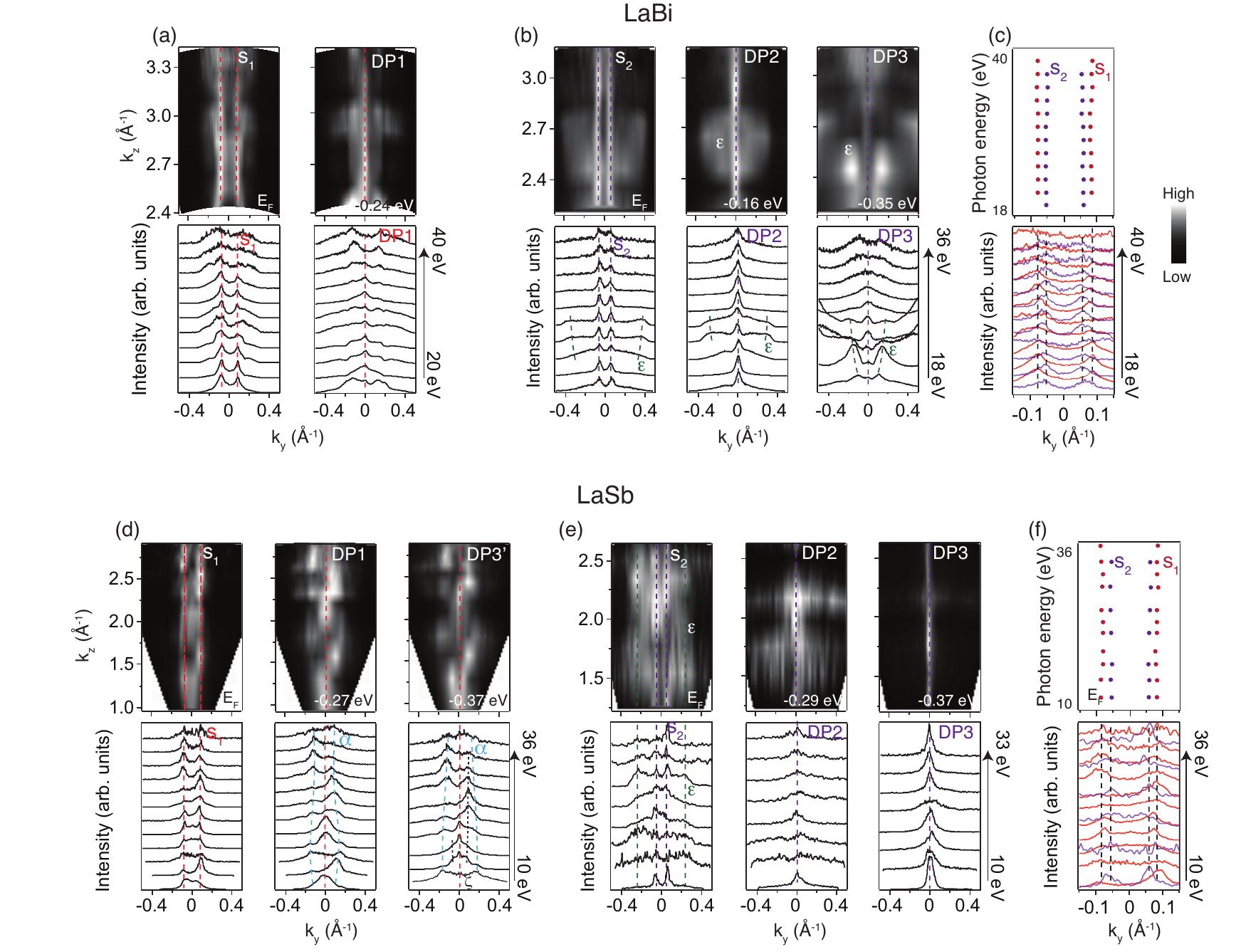}
\caption {\label{kz} The Fermi surface and photoemission intensity below $E_F$ plotted as a function of $k_z$ for LaBi and LaSb. Photoemission intensity maps and corresponding Momentum Distribution Curves (MDCs) around the surface Brillouin zone (a) center and (b) corner in the $k_y$-$k_z$ plane of LaBi are shown at at $E_F$ and at the Dirac points' binding energies. (c) Fermi crossings of the $s_1$ and $s_2$ bands of LaBi as a function of photon energy. (d), (e) $k_y$-$k_z$-plane maps and corresponding MDCs of LaSb. (f) Fermi crossings of the $s_1$ and $s_2$ bands of LaSb as a function of photon energy. The dashed lines show the bands and Dirac points, where identifiable. Different $k_z$s were accessed by varying the photon energy.}
\end{figure*}

\begin{figure*}[htb]
\includegraphics[width=\textwidth]{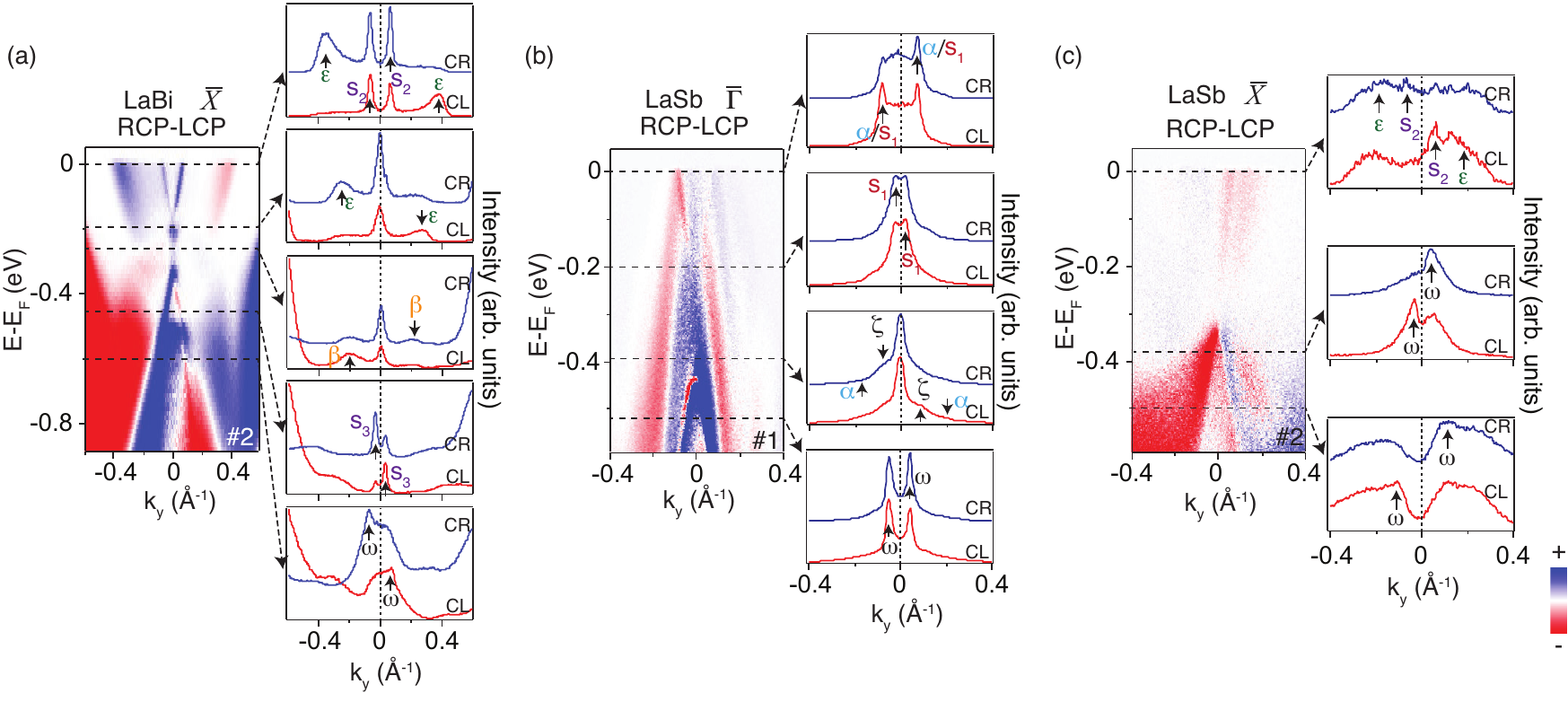}
\caption {\label{cd} Circular dichroism of band structures at $\bar{\varGamma}$ and $\bar{X}$. (a) The difference intensity plot of cut \#1 in Fig.~\ref{FS}(e) of LaBi measured under RCP and LCP light, with several illustrative MDCs at binding energies and with polarizations as marked. (b, c) Similar data on LaSb on cuts \#1 and \#2 in Fig.~\ref{FS}(g), respectively. Data in panels a, b and c were taken with 22, 22 and 26~eV photons, respectively.}
\end{figure*}

\begin{figure}[htb] 
\includegraphics[width=\columnwidth]{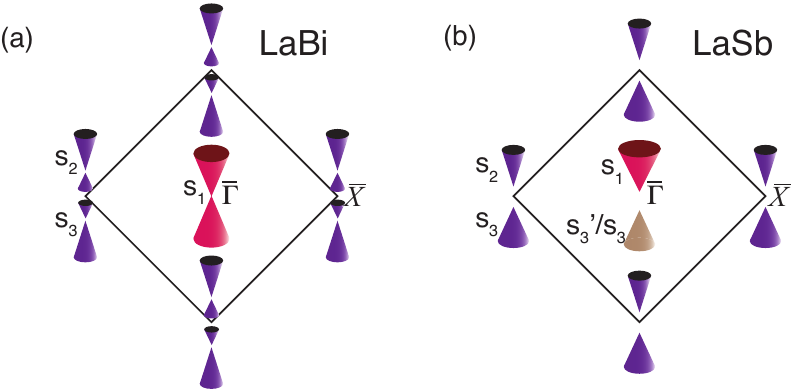}
\caption{\label{DP} Illustrations of the Dirac cones in the surface Brillouin zones of LaBi and LaSb. LaBi has one Dirac cone at $\bar{\varGamma}$ and two Dirac cones at $\bar{X}$.  LaSb has two Dirac cones at $\bar{X}$ and one or two Dirac cones at $\bar{\varGamma}$. The light brown Dirac cone may be a projection from the $X_3$ point.
}
\end{figure}

\begin{figure}[htb] 
\includegraphics[width=\columnwidth]{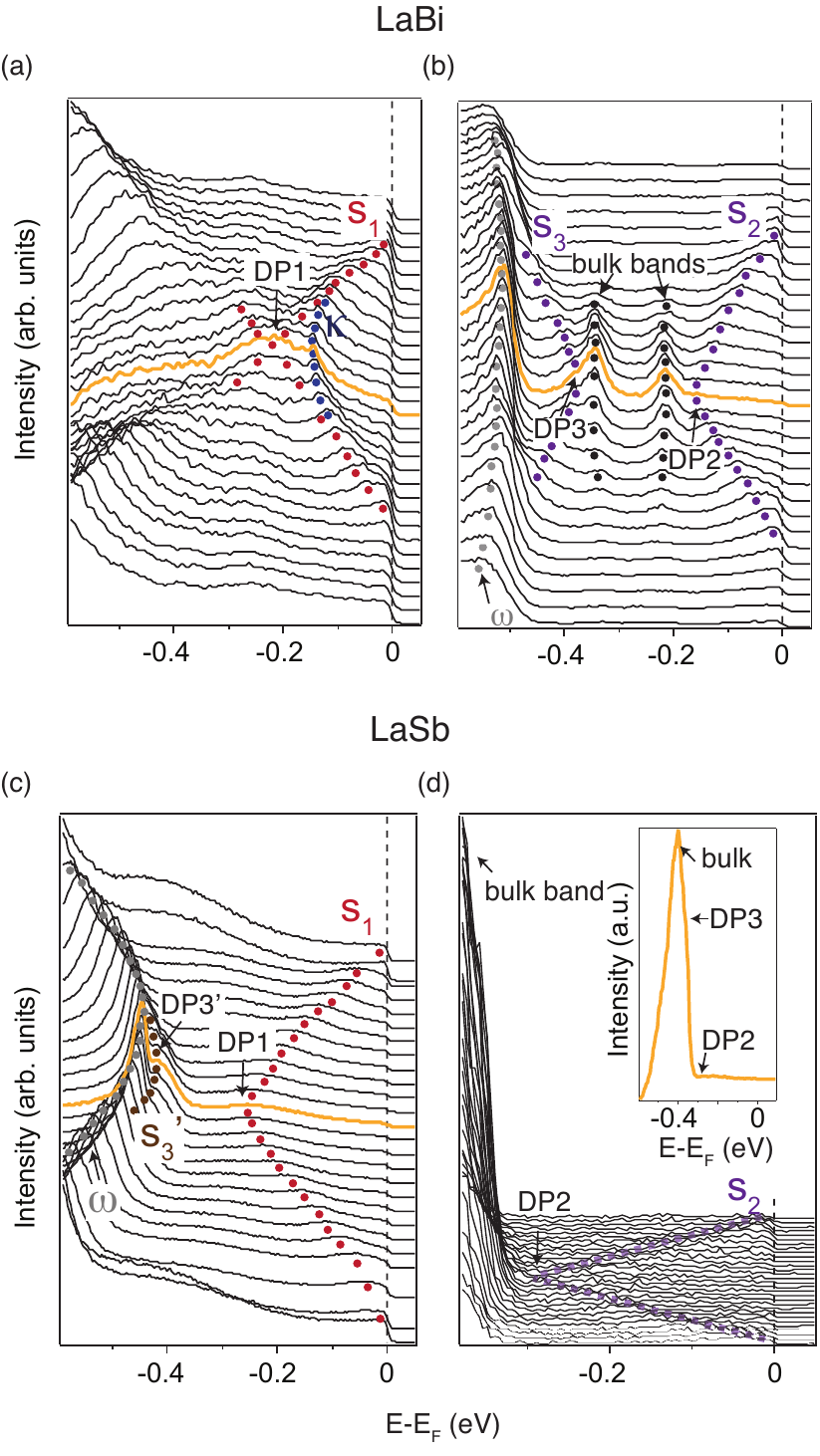}
\caption{\label{EDC} (a,b) and (c,d) The EDCs for the photoemission intensity in Figs.~\ref{bands}(a,b) and (d,e), respectively. The orange curves identify the EDCs at the $\bar{\varGamma}$ and $\bar{X}$ points.}
\end{figure}


The band dispersions of LaBi and LaSb are shown in Fig.~\ref{bands}.  Besides the calculated bulk bands\cite{Guo2016, Kumigashira1998}, we observe several surface bands. Figs.~\ref{bands}(a) and (b), respectively, show the photoemission intensity of LaBi along cuts \#1 and 2 together with the corresponding 2D curvature intensity plots. Around the $\bar{\varGamma}$ point in Fig.~\ref{bands}(a), we are able to resolve two hole-like bands ($\alpha$ and $\beta$), one electron-like band ($\gamma$), and surface $s_1$ band which has a linear band dispersion and an apparent Dirac point at a binding energy of $-$0.24\,eV. The $\alpha$ and $\beta$ bands are consistent with band structure calculations\cite{Guo2016}, while the parabolic-like $\gamma$ band is the projection of the electron pocket at the $X_3$ point. The bottom of the $\gamma$ band and the Dirac point (DP1) of the surface $s_1$ band can be clearly distinguished. 
In Fig.~\ref{bands}(b), there is a clear anti-crossing of the $\beta$ and $\varepsilon$ bands along the $\bar{\varGamma}$-$\bar{X}$ direction, which is predicted by band structure calculations\cite{Fuliang2015} but was not observed in the previous ARPES data\cite{Kumigashira1998,Wu2016,Nayak2016}. Interestingly, there are two apparent Dirac points at the $\bar{X}$ point --- DP2 and DP3 occur at binding energies of $-$0.16 and $-$0.35\,eV, respectively. The bulk band $\delta$ is too weak to be observed at $\bar{X}$, possibly due to the strong signal from the surface $s_2$ band. At cut \#3 near the $\bar{X}$ point [Fig.~\ref{bands}(c)], we can clearly see both the parabolic $\delta$ band and the surface $s_2$ band. 

For LaSb, around $\bar{\varGamma}$ in Fig.~\ref{bands}(d), there are two hole-like bands ($\alpha$ and $\beta$) and a linear surface band $s_1$ with its Dirac point DP1 located at $-$0.27\,eV. The black dashed line below $-$0.4\,eV indicates the parabolic bulk $\omega$ band. Above the $ \omega $ band, there is a second apparent Dirac cone DP3', with its apex at $-$0.37\,eV. The $\gamma$ band is not clearly resolved in cut \#1.  The bands near $E_F$ around $\bar{X}$ are much weaker in Fig.~\ref{bands}(e), preventing observation of the band anti-crossing along the $\bar{\varGamma}$-$\bar{X}$ direction. However, similar to the surface states in LaBi, the linearly-dispersing $s_2$ and $s_3$ surface states and the two Dirac cones (DP2 at $-$0.29\,eV and DP3 at $-$0.37\,eV) still exist at $\bar{X}$ in LaSb.  However, only halves of the Dirac-cone-like bands can be seen in cut~\#1 and \#2.
Fig.~\ref{bands}(f) is the band structure on cut~\#3 along the $\bar{\varGamma}$-$\bar{M}$ direction. Here we can see the parabolic-like bulk $\gamma$ band more clearly than in Fig.~\ref{bands}(d), demonstrating the close similarity to LaBi. An additional hole-like band $\zeta$ is observed which can be found in the bulk band calculations\cite{Kumigashira1998}.

To verify the two-dimensional nature of the surface states, we performed photon-energy-dependent measurements from 10 to 40\,eV to scan $k_z$. Figs.~\ref{kz}(a) and (b) show the $k_z$-$k_y$-plane maps of LaBi at $E_F$ and selected higher binding energies around the BZ center and corner, respectively. The $s_1$ and $s_2$ bands show two-dimensional character. The $k_z$-$k_y$-plane maps of DP1, DP2 and DP3 demonstrate that these states do not disperse along the $k_z$ direction, identifying them as surface states. Some of the intensity in the $k_z$-$k_y$-plane maps derives from the bulk bands.
The $k_z$ dispersions of the Fermi crossings of $s_1$ and $s_2$ in Fig.~\ref{kz}(c) demonstrate that these are two totally distinct, $k_z$-nondispersive surface bands, while the three Dirac points occur at different energies. Thus the $s_1$ band cannot be a projection from the $X_3$ point. 
Figs.~\ref{kz}(d) and (e) show similar $k_z$-$k_y$-plane maps for LaSb. The $s_1$ and $s_2$ bands and the Dirac points show two-dimensional bahavior as in LaBi. The $k_z$ dispersions of the Fermi crossings of $s_1$ and $s_2$ in Fig.~\ref{kz}(f) also demonstrate that the $s_1$ band cannot be a projection of the $s_2$ band from the $X_3$ point. 
However, a similar comparison is difficult for the $s_3'$ and $s_3$ bands in LaSb because of their proximity to bulk bands. Dirac points DP3' and DP3 occur at the same energy ($-$0.37\,eV), suggesting that the $s_3'$ band is the projection of the $s_3$ band from the $X_3$ point, but this remains to be confirmed. If the $s_3'$ band is indeed the projection of the $s_3$ band, the total number of Dirac points below $E_F$ in both compounds would be three, which would suggest that they are both topologically nontrivial.

The differential coupling of right-circularly (RCP) versus left-circularly (LCP) polarized light can reveal the orbital angular momentum (OAM) of an electronic state\cite{Wang2013CD}. The difference between transition matrix elements of photoelectron final states results in circular dichroism (CD) proportional to the inner product of the OAM direction and the incoming photon direction\cite{Park2012CD}. In strong spin-orbit-coupled materials, both the spin and orbital angular momenta of a state would exhibit conjugate textures around the Fermi surface. Therefore, to further explore the spin texture of the surface states, we performed CD-ARPES experiments.

Fig.~\ref{cd} shows the CD of the ARPES band structures and momentum distribution curves (MDCs) at illustrative binding energies under RCP and LCP light. The photoemission intensity exhibits strong intensity inversion between RCP and LCP data. Interestingly, from the detailed MDCs, we can see clear CD of both the surface and bulk bands. Circular dichroism is natural and expected for topological surface states, and supports the ``forbidden backscattering'' mechanism for magnetoresistance of the surface state. However, bulk La$X$ materials are believed to be both inversion- and time-reversal-symmetric. In such a case, one can strictly prove the spin must be degenerate\cite{Dresselhaus1955,fu2007TI2}, regardless of SOC effects; therefore, the ``forbidden backscattering'' mechanism should not apply to the bulk bands. However, ARPES probes the near-surface electronic structure where inversion is explicitly broken by the presence of a surface. The near-surface bulk band structure detected at the surface can vary from the actual bulk band structure. Thus the CD observed for the near-surface bulk bands may suggest that these bands are strongly influenced by the surface, and forbidden backscattering may play a role in the magnetoresistance for all states near the surface, not just the Dirac cones.  This may be similar to the case of WTe$_2$\cite{Jiang2015} and Cd$_3$As$_2$\cite{liang2015CdAs}.  

\section{Discussion}


\begin{table}[htb]
  \caption{\label{carrier density} ARPES Fermi surface volumes and carrier densities, compared with those from de~Haas-van~Alphen (dHvA). Labelling of Fermi surface sheets refers to Fig.~\ref{FS}(e); dHvA values were extracted from Ref.~\onlinecite{Hasegawa1985}, based on data in Ref.~\onlinecite{Kitazawa1983}.}
  \begin{tabular}{lclcccc}\hline
    \multicolumn{2}{c}{Sheet} & Type & \multicolumn{2}{c}{Volume (nm$^{-3}$)} & \multicolumn{2}{c}{Carrier density (cm$^{-3}$)} \\
    & & & ARPES & dHvA & ARPES & dHvA \\ \hline\hline
      &$\alpha$ & hole, $\varGamma$ & 9.6 & 11.1 & 7.75$\times$10$^{19}$ & 8.98$\times$10$^{19}$\\
     LaBi &$\beta$  &hole, $\varGamma$  & 22.2 & 38.2 & 1.79$\times$10$^{20}$ & 3.09$\times$10$^{20}$\\
      & $\delta/\varepsilon$ & electron, $X$ & 30.3 & 46.9 & 2.45$\times$10$^{20}$ & 3.79$\times$10$^{20}$\\  \hline
      & $\alpha$ & hole, $\varGamma$ & 3.4 & 7.6 & 2.72$\times$10$^{19}$ & 6.01$\times$10$^{19}$\\
     LaSb & $\beta$ & hole, $\varGamma$  & 14.8 & 20.8 & 1.19$\times$10$^{20}$ & 1.68$\times$10$^{20}$\\
      & $\delta/\varepsilon$ & electron, $X$ & 15.0 & 26.4 & 1.21$\times$10$^{20}$ & 2.14$\times$10$^{20}$\\ \hline
  \end{tabular}
\end{table}

Although the near-surface bulk bands observed by ARPES may vary from the actual bulk bands, it is worth comparing the ARPES-derived Fermi surfaces with those found by quantum oscillations\cite{Kitazawa1983,Settai1993}. The two-dimensional projected Fermi surfaces in Fig.~\ref{FS}(d, f) allow us to estimate the Fermi pocket volumes by also considering the calculated Fermi pockets' topology and noting the cubic symmetry. We regard the two-dimensional Fermi surfaces as the maximum section of the three-dimensional Fermi pockets. The electron and hole Fermi pocket volumes from ARPES are roughly half those from dHvA. Since our quantum oscillation data on the same crystals are consistent with dHvA, the discrepancy is presumably due to the influence of the surface on the ARPES spectra. Our bulk-like Fermi surfaces and band structures do contain the essential features found in the calculations, even if the bands are shifted.  In Tab.~\ref{carrier density}, we also calculated the carrier densities from our ARPES data. The ratio of the populations of the holes and electrons is about 1.05 and 1.21 for LaBi and LaSb, respectively, i.e.\ electrons and holes are compensated within our experimental error bar. Although the surface carrier is just a small fraction of the bulk, because of the back scattering, its conductivity can be much higher than the bulk, so its contribution to MR should be noted. Thus our findings suggest that the XMR may result from the carrier compensation in addition to the ``forbidden backscattering'' mechanism of surface and near-surface bulk states.

The surface band structure in LaBi includes three clear, fully-resolved Dirac points, although DP2 and DP3 are set off by only a few tens of meV from the bulk bands between them.  LaSb, however, shows three or possibly four Dirac cones seemingly terminating at Dirac points, likely due to broadening from proximity to the bulk bands. This situation is summarized schematically in Fig.~\ref{DP}, with further detail provided in Figs.~\ref{EDC}(a-d).  Band structure calculations with spin-orbit coupling indicate band inversion at the $X$ point between the La $d$ states and the pnictogen $p$ states, with an anti-crossing along the $\varGamma$-$X$ direction\cite{Zeng2016}. The gap we observe at the anti-crossing in LaBi is 20\,meV [see Fig.~\ref{bands}(b)], which is smaller than the calculated 35\,meV\cite{Fuliang2015}. The reduced spin-orbit coupling in lighter Sb is most likely responsible for the reduced separation of the Dirac states from the bulk bands compared with the Bi material.  The bulk band structure observed in LaBi fits well with the calculations, while the intervening bulk band is not observed in LaSb.  However, the authors are not aware of a slab calculation relevant to (001) surfaces in these materials.  To fully interpret the data, further band calculations are required beyond those in Refs.~\onlinecite{Fuliang2015,Kumar2016}.

It is worth comparing the present results with recent ARPES measurements on LaBi and LaSb. In the case of LaBi, Ref.~\onlinecite{Wu2016} reports both surface and bulk bands at $\bar{\varGamma}$; however, the $s_3'$ band was not resolved and data at $\bar{X}$ are not reported. Ref.~\onlinecite{Nayak2016} presents linear surface bands at both $\bar{\varGamma}$ and $\bar{X}$, similar to the present work, but the two distinct Dirac cones at the $\bar{X}$ point are not clearly resolved. Furthermore, the band anti-crossing shown in Fig.~\ref{bands}(b) is not visible in Refs.~\onlinecite{Wu2016, Nayak2016}. A consistent picture emerges for this material, with higher-resolution data and circular dichroism providing crucial additional detail. Our data on LaSb are similar to those in Ref.~\onlinecite{Zeng2016}, the primary difference being that Zeng $et~al$ did not identify the surface bands. Hints of the Dirac cones are indeed visible in their $\bar{X}$-point data, but the proximity to faint bulk bands makes these features more difficult to discern.  While circular dichroism makes a strong case for the existence of spin-polarized surface states, and it would be unlikely that the reduction in spin-orbit coupling from Bi to Sb would eliminate these features entirely, the different interpretations possible based on these data would be best resolved through detailed slab calculations.  

Finally, it is worth commenting further on the quantum oscillations:  Shubnikov-de~Haas oscillations were readily observed to remarkably low fields at 1.8\,K, without the use of specialized apparatus, in all samples measured, and without any optimization of the crystal growth beyond its initial success.  This implies that these systems are not just crystallographically simple, but also extremely clean and easy to prepare, and suggests that they may be an excellent system for future in-depth study.

\section{Conclusion}
In summary, we have observed the surface bands and near-surface bulk electronic states on the (001) surface of LaBi and LaSb using ARPES, and identify an obvious band anti-crossing along the $\bar{\varGamma}$-$\bar{X}$ direction in LaBi. An odd number of Dirac cones are clearly present below the Fermi level in LaBi. Dirac surface states are also observed in LaSb, but the Dirac points are much closer to the bulk bands, and we cannot be certain whether an odd or even number lie below $E_F$. Furthermore, strong circular dichroism indicates spin-orbit coupling-induced spin and orbital angular momentum textures in both the surface and near-surface bulk bands, which likely contribute to the large magnetoresistance in addition to the electron-hole compensation. Our high-resolution electronic structures of LaBi and LaSb lay the foundation for further investigations. Since preparing excellent crystals of both materials is relatively straightforward and the crystal structure is extremely simple, the lanthanum pnictide family may provide a uniquely accessible experimental platform for investigating topological surface states and their evolution as spin-orbit coupling increases down the periodic table.

\section*{Acknowledgments}
Some preliminary data (not shown here) were taken at beamline 13U of the National Synchrotron Radiation Laboratory (NSRL) and beamline 09U of the Shanghai Synchrotron Radiation Facility (SSRF). We gratefully acknowledge helpful discussions with Professor X. G. Wan of Nanjing University and experimental support by Dr.\ D.\ H.\ Lu and Dr.\ H.\ Makoto at SSRL. This work is supported by the National Key R$\&$D Program of the MOST of China (Grant No.~2016YFA0300203) and the Science Challenge Program of China. Use of the Stanford Synchrotron Radiation Lightsource, SLAC National Accelerator Laboratory, is supported by the U.S.\ Department of Energy, Office of Science, Office of Basic Energy Sciences under Contract No.\ DE-AC02-76SF00515.

\bibliography{LaX_prb}

\begin{thebibliography}{51}%
\makeatletter
\providecommand \@ifxundefined [1]{%
 \@ifx{#1\undefined}
}%
\providecommand \@ifnum [1]{%
 \ifnum #1\expandafter \@firstoftwo
 \else \expandafter \@secondoftwo
 \fi
}%
\providecommand \@ifx [1]{%
 \ifx #1\expandafter \@firstoftwo
 \else \expandafter \@secondoftwo
 \fi
}%
\providecommand \natexlab [1]{#1}%
\providecommand \enquote  [1]{``#1''}%
\providecommand \bibnamefont  [1]{#1}%
\providecommand \bibfnamefont [1]{#1}%
\providecommand \citenamefont [1]{#1}%
\providecommand \href@noop [0]{\@secondoftwo}%
\providecommand \href [0]{\begingroup \@sanitize@url \@href}%
\providecommand \@href[1]{\@@startlink{#1}\@@href}%
\providecommand \@@href[1]{\endgroup#1\@@endlink}%
\providecommand \@sanitize@url [0]{\catcode `\\12\catcode `\$12\catcode
  `\&12\catcode `\#12\catcode `\^12\catcode `\_12\catcode `\%12\relax}%
\providecommand \@@startlink[1]{}%
\providecommand \@@endlink[0]{}%
\providecommand \url  [0]{\begingroup\@sanitize@url \@url }%
\providecommand \@url [1]{\endgroup\@href {#1}{\urlprefix }}%
\providecommand \urlprefix  [0]{URL }%
\providecommand \Eprint [0]{\href }%
\providecommand \doibase [0]{http://dx.doi.org/}%
\providecommand \selectlanguage [0]{\@gobble}%
\providecommand \bibinfo  [0]{\@secondoftwo}%
\providecommand \bibfield  [0]{\@secondoftwo}%
\providecommand \translation [1]{[#1]}%
\providecommand \BibitemOpen [0]{}%
\providecommand \bibitemStop [0]{}%
\providecommand \bibitemNoStop [0]{.\EOS\space}%
\providecommand \EOS [0]{\spacefactor3000\relax}%
\providecommand \BibitemShut  [1]{\csname bibitem#1\endcsname}%
\let\auto@bib@innerbib\@empty
\bibitem [{\citenamefont {Bernevig}\ \emph {et~al.}(2006)\citenamefont
  {Bernevig}, \citenamefont {Hughes},\ and\ \citenamefont
  {Zhang}}]{bernevig2006TI}%
  \BibitemOpen
  \bibfield  {author} {\bibinfo {author} {\bibfnamefont {B.~A.}\ \bibnamefont
  {Bernevig}}, \bibinfo {author} {\bibfnamefont {T.~L.}\ \bibnamefont
  {Hughes}}, \ and\ \bibinfo {author} {\bibfnamefont {S.-C.}\ \bibnamefont
  {Zhang}},\ }\href {\doibase 10.1126/science.1133734} {\bibfield  {journal}
  {\bibinfo  {journal} {Science}\ }\textbf {\bibinfo {volume} {314}},\ \bibinfo
  {pages} {1757} (\bibinfo {year} {2006})}\BibitemShut {NoStop}%
\bibitem [{\citenamefont {Zhang}\ \emph {et~al.}(2009)\citenamefont {Zhang},
  \citenamefont {Liu}, \citenamefont {Qi}, \citenamefont {Dai}, \citenamefont
  {Fang},\ and\ \citenamefont {Zhang}}]{zhang2009TI}%
  \BibitemOpen
  \bibfield  {author} {\bibinfo {author} {\bibfnamefont {H.}~\bibnamefont
  {Zhang}}, \bibinfo {author} {\bibfnamefont {C.-X.}\ \bibnamefont {Liu}},
  \bibinfo {author} {\bibfnamefont {X.-L.}\ \bibnamefont {Qi}}, \bibinfo
  {author} {\bibfnamefont {X.}~\bibnamefont {Dai}}, \bibinfo {author}
  {\bibfnamefont {Z.}~\bibnamefont {Fang}}, \ and\ \bibinfo {author}
  {\bibfnamefont {S.-C.}\ \bibnamefont {Zhang}},\ }\href {\doibase
  10.1038/nphys1270} {\bibfield  {journal} {\bibinfo  {journal} {Nature
  physics}\ }\textbf {\bibinfo {volume} {5}},\ \bibinfo {pages} {438} (\bibinfo
  {year} {2009})}\BibitemShut {NoStop}%
\bibitem [{\citenamefont {Xia}\ \emph {et~al.}(2009)\citenamefont {Xia},
  \citenamefont {Qian}, \citenamefont {Hsieh}, \citenamefont {Wray},
  \citenamefont {Pal}, \citenamefont {Lin}, \citenamefont {Bansil},
  \citenamefont {Grauer}, \citenamefont {Hor}, \citenamefont {Cava},\ and\
  \citenamefont {Hasan}}]{xia2009TI}%
  \BibitemOpen
  \bibfield  {author} {\bibinfo {author} {\bibfnamefont {Y.}~\bibnamefont
  {Xia}}, \bibinfo {author} {\bibfnamefont {D.}~\bibnamefont {Qian}}, \bibinfo
  {author} {\bibfnamefont {D.}~\bibnamefont {Hsieh}}, \bibinfo {author}
  {\bibfnamefont {L.}~\bibnamefont {Wray}}, \bibinfo {author} {\bibfnamefont
  {A.}~\bibnamefont {Pal}}, \bibinfo {author} {\bibfnamefont {H.}~\bibnamefont
  {Lin}}, \bibinfo {author} {\bibfnamefont {A.}~\bibnamefont {Bansil}},
  \bibinfo {author} {\bibfnamefont {D.}~\bibnamefont {Grauer}}, \bibinfo
  {author} {\bibfnamefont {Y.}~\bibnamefont {Hor}}, \bibinfo {author}
  {\bibfnamefont {R.}~\bibnamefont {Cava}}, \ and\ \bibinfo {author}
  {\bibfnamefont {M.}~\bibnamefont {Hasan}},\ }\href {\doibase
  10.1038/nphys1274} {\bibfield  {journal} {\bibinfo  {journal} {Nature
  Physics}\ }\textbf {\bibinfo {volume} {5}},\ \bibinfo {pages} {398} (\bibinfo
  {year} {2009})}\BibitemShut {NoStop}%
\bibitem [{\citenamefont {Fu}\ \emph {et~al.}(2007)\citenamefont {Fu},
  \citenamefont {Kane},\ and\ \citenamefont {Mele}}]{fu2007TI}%
  \BibitemOpen
  \bibfield  {author} {\bibinfo {author} {\bibfnamefont {L.}~\bibnamefont
  {Fu}}, \bibinfo {author} {\bibfnamefont {C.~L.}\ \bibnamefont {Kane}}, \ and\
  \bibinfo {author} {\bibfnamefont {E.~J.}\ \bibnamefont {Mele}},\ }\href
  {\doibase 10.1103/PhysRevLett.98.106803} {\bibfield  {journal} {\bibinfo
  {journal} {Phys. Rev. Lett.}\ }\textbf {\bibinfo {volume} {98}},\ \bibinfo
  {pages} {106803} (\bibinfo {year} {2007})}\BibitemShut {NoStop}%
\bibitem [{\citenamefont {Fu}\ and\ \citenamefont {Kane}(2007)}]{fu2007TI2}%
  \BibitemOpen
  \bibfield  {author} {\bibinfo {author} {\bibfnamefont {L.}~\bibnamefont
  {Fu}}\ and\ \bibinfo {author} {\bibfnamefont {C.~L.}\ \bibnamefont {Kane}},\
  }\href {\doibase 10.1103/PhysRevB.76.045302} {\bibfield  {journal} {\bibinfo
  {journal} {Phys. Rev. B}\ }\textbf {\bibinfo {volume} {76}},\ \bibinfo
  {pages} {045302} (\bibinfo {year} {2007})}\BibitemShut {NoStop}%
\bibitem [{\citenamefont {Teo}\ \emph {et~al.}(2008)\citenamefont {Teo},
  \citenamefont {Fu},\ and\ \citenamefont {Kane}}]{Teo2008TI}%
  \BibitemOpen
  \bibfield  {author} {\bibinfo {author} {\bibfnamefont {J.~C.~Y.}\
  \bibnamefont {Teo}}, \bibinfo {author} {\bibfnamefont {L.}~\bibnamefont
  {Fu}}, \ and\ \bibinfo {author} {\bibfnamefont {C.~L.}\ \bibnamefont
  {Kane}},\ }\href {\doibase 10.1103/PhysRevB.78.045426} {\bibfield  {journal}
  {\bibinfo  {journal} {Phys. Rev. B}\ }\textbf {\bibinfo {volume} {78}},\
  \bibinfo {pages} {045426} (\bibinfo {year} {2008})}\BibitemShut {NoStop}%
\bibitem [{\citenamefont {Hsieh}\ \emph {et~al.}(2008)\citenamefont {Hsieh},
  \citenamefont {Qian}, \citenamefont {Wray}, \citenamefont {Xia},
  \citenamefont {Hor}, \citenamefont {Cava},\ and\ \citenamefont
  {Hasan}}]{hsieh2008TI}%
  \BibitemOpen
  \bibfield  {author} {\bibinfo {author} {\bibfnamefont {D.}~\bibnamefont
  {Hsieh}}, \bibinfo {author} {\bibfnamefont {D.}~\bibnamefont {Qian}},
  \bibinfo {author} {\bibfnamefont {L.}~\bibnamefont {Wray}}, \bibinfo {author}
  {\bibfnamefont {Y.}~\bibnamefont {Xia}}, \bibinfo {author} {\bibfnamefont
  {Y.~S.}\ \bibnamefont {Hor}}, \bibinfo {author} {\bibfnamefont
  {R.}~\bibnamefont {Cava}}, \ and\ \bibinfo {author} {\bibfnamefont {M.~Z.}\
  \bibnamefont {Hasan}},\ }\href {\doibase 10.1038/nature06843} {\bibfield
  {journal} {\bibinfo  {journal} {Nature}\ }\textbf {\bibinfo {volume} {452}},\
  \bibinfo {pages} {970} (\bibinfo {year} {2008})}\BibitemShut {NoStop}%
\bibitem [{\citenamefont {Chen}\ \emph {et~al.}(2009)\citenamefont {Chen},
  \citenamefont {Analytis}, \citenamefont {Chu}, \citenamefont {Liu},
  \citenamefont {Mo}, \citenamefont {Qi}, \citenamefont {Zhang}, \citenamefont
  {Lu}, \citenamefont {Dai}, \citenamefont {Fang} \emph
  {et~al.}}]{chen2009BiTe}%
  \BibitemOpen
  \bibfield  {author} {\bibinfo {author} {\bibfnamefont {Y.}~\bibnamefont
  {Chen}}, \bibinfo {author} {\bibfnamefont {J.}~\bibnamefont {Analytis}},
  \bibinfo {author} {\bibfnamefont {J.-H.}\ \bibnamefont {Chu}}, \bibinfo
  {author} {\bibfnamefont {Z.}~\bibnamefont {Liu}}, \bibinfo {author}
  {\bibfnamefont {S.-K.}\ \bibnamefont {Mo}}, \bibinfo {author} {\bibfnamefont
  {X.-L.}\ \bibnamefont {Qi}}, \bibinfo {author} {\bibfnamefont
  {H.}~\bibnamefont {Zhang}}, \bibinfo {author} {\bibfnamefont
  {D.}~\bibnamefont {Lu}}, \bibinfo {author} {\bibfnamefont {X.}~\bibnamefont
  {Dai}}, \bibinfo {author} {\bibfnamefont {Z.}~\bibnamefont {Fang}},  \emph
  {et~al.},\ }\href {\doibase 10.1126/science.1173034} {\bibfield  {journal}
  {\bibinfo  {journal} {Science}\ }\textbf {\bibinfo {volume} {325}},\ \bibinfo
  {pages} {178} (\bibinfo {year} {2009})}\BibitemShut {NoStop}%
\bibitem [{\citenamefont {Young}\ \emph {et~al.}(2012)\citenamefont {Young},
  \citenamefont {Zaheer}, \citenamefont {Teo}, \citenamefont {Kane},
  \citenamefont {Mele},\ and\ \citenamefont {Rappe}}]{Young2012dirac}%
  \BibitemOpen
  \bibfield  {author} {\bibinfo {author} {\bibfnamefont {S.~M.}\ \bibnamefont
  {Young}}, \bibinfo {author} {\bibfnamefont {S.}~\bibnamefont {Zaheer}},
  \bibinfo {author} {\bibfnamefont {J.~C.~Y.}\ \bibnamefont {Teo}}, \bibinfo
  {author} {\bibfnamefont {C.~L.}\ \bibnamefont {Kane}}, \bibinfo {author}
  {\bibfnamefont {E.~J.}\ \bibnamefont {Mele}}, \ and\ \bibinfo {author}
  {\bibfnamefont {A.~M.}\ \bibnamefont {Rappe}},\ }\href {\doibase
  10.1103/PhysRevLett.108.140405} {\bibfield  {journal} {\bibinfo  {journal}
  {Phys. Rev. Lett.}\ }\textbf {\bibinfo {volume} {108}},\ \bibinfo {pages}
  {140405} (\bibinfo {year} {2012})}\BibitemShut {NoStop}%
\bibitem [{\citenamefont {Liu}\ \emph {et~al.}(2014{\natexlab{a}})\citenamefont
  {Liu}, \citenamefont {Zhou}, \citenamefont {Zhang}, \citenamefont {Wang},
  \citenamefont {Weng}, \citenamefont {Prabhakaran}, \citenamefont {Mo},
  \citenamefont {Shen}, \citenamefont {Fang}, \citenamefont {Dai} \emph
  {et~al.}}]{Liu2014Na3Bi}%
  \BibitemOpen
  \bibfield  {author} {\bibinfo {author} {\bibfnamefont {Z.}~\bibnamefont
  {Liu}}, \bibinfo {author} {\bibfnamefont {B.}~\bibnamefont {Zhou}}, \bibinfo
  {author} {\bibfnamefont {Y.}~\bibnamefont {Zhang}}, \bibinfo {author}
  {\bibfnamefont {Z.}~\bibnamefont {Wang}}, \bibinfo {author} {\bibfnamefont
  {H.}~\bibnamefont {Weng}}, \bibinfo {author} {\bibfnamefont {D.}~\bibnamefont
  {Prabhakaran}}, \bibinfo {author} {\bibfnamefont {S.-K.}\ \bibnamefont {Mo}},
  \bibinfo {author} {\bibfnamefont {Z.}~\bibnamefont {Shen}}, \bibinfo {author}
  {\bibfnamefont {Z.}~\bibnamefont {Fang}}, \bibinfo {author} {\bibfnamefont
  {X.}~\bibnamefont {Dai}},  \emph {et~al.},\ }\href {\doibase
  10.1126/science.1245085} {\bibfield  {journal} {\bibinfo  {journal}
  {Science}\ }\textbf {\bibinfo {volume} {343}},\ \bibinfo {pages} {864}
  (\bibinfo {year} {2014}{\natexlab{a}})}\BibitemShut {NoStop}%
\bibitem [{\citenamefont {Liu}\ \emph {et~al.}(2014{\natexlab{b}})\citenamefont
  {Liu}, \citenamefont {Jiang}, \citenamefont {Zhou}, \citenamefont {Wang},
  \citenamefont {Zhang}, \citenamefont {Weng}, \citenamefont {Prabhakaran},
  \citenamefont {Mo}, \citenamefont {Peng}, \citenamefont {Dudin} \emph
  {et~al.}}]{Liu2014Cd3As2}%
  \BibitemOpen
  \bibfield  {author} {\bibinfo {author} {\bibfnamefont {Z.}~\bibnamefont
  {Liu}}, \bibinfo {author} {\bibfnamefont {J.}~\bibnamefont {Jiang}}, \bibinfo
  {author} {\bibfnamefont {B.}~\bibnamefont {Zhou}}, \bibinfo {author}
  {\bibfnamefont {Z.}~\bibnamefont {Wang}}, \bibinfo {author} {\bibfnamefont
  {Y.}~\bibnamefont {Zhang}}, \bibinfo {author} {\bibfnamefont
  {H.}~\bibnamefont {Weng}}, \bibinfo {author} {\bibfnamefont {D.}~\bibnamefont
  {Prabhakaran}}, \bibinfo {author} {\bibfnamefont {S.}~\bibnamefont {Mo}},
  \bibinfo {author} {\bibfnamefont {H.}~\bibnamefont {Peng}}, \bibinfo {author}
  {\bibfnamefont {P.}~\bibnamefont {Dudin}},  \emph {et~al.},\ }\href {\doibase
  10.1038/nmat3990} {\bibfield  {journal} {\bibinfo  {journal} {Nature
  materials}\ }\textbf {\bibinfo {volume} {13}},\ \bibinfo {pages} {677}
  (\bibinfo {year} {2014}{\natexlab{b}})}\BibitemShut {NoStop}%
\bibitem [{\citenamefont {Xu}\ \emph {et~al.}(2015{\natexlab{a}})\citenamefont
  {Xu}, \citenamefont {Liu}, \citenamefont {Kushwaha}, \citenamefont {Sankar},
  \citenamefont {Krizan}, \citenamefont {Belopolski}, \citenamefont {Neupane},
  \citenamefont {Bian}, \citenamefont {Alidoust}, \citenamefont {Chang} \emph
  {et~al.}}]{Xu2015dirac}%
  \BibitemOpen
  \bibfield  {author} {\bibinfo {author} {\bibfnamefont {S.-Y.}\ \bibnamefont
  {Xu}}, \bibinfo {author} {\bibfnamefont {C.}~\bibnamefont {Liu}}, \bibinfo
  {author} {\bibfnamefont {S.~K.}\ \bibnamefont {Kushwaha}}, \bibinfo {author}
  {\bibfnamefont {R.}~\bibnamefont {Sankar}}, \bibinfo {author} {\bibfnamefont
  {J.~W.}\ \bibnamefont {Krizan}}, \bibinfo {author} {\bibfnamefont
  {I.}~\bibnamefont {Belopolski}}, \bibinfo {author} {\bibfnamefont
  {M.}~\bibnamefont {Neupane}}, \bibinfo {author} {\bibfnamefont
  {G.}~\bibnamefont {Bian}}, \bibinfo {author} {\bibfnamefont {N.}~\bibnamefont
  {Alidoust}}, \bibinfo {author} {\bibfnamefont {T.-R.}\ \bibnamefont {Chang}},
   \emph {et~al.},\ }\href {\doibase 10.1126/science.1256742} {\bibfield
  {journal} {\bibinfo  {journal} {Science}\ }\textbf {\bibinfo {volume}
  {347}},\ \bibinfo {pages} {294} (\bibinfo {year}
  {2015}{\natexlab{a}})}\BibitemShut {NoStop}%
\bibitem [{\citenamefont {Wan}\ \emph {et~al.}(2011)\citenamefont {Wan},
  \citenamefont {Turner}, \citenamefont {Vishwanath},\ and\ \citenamefont
  {Savrasov}}]{Wan2011weyl}%
  \BibitemOpen
  \bibfield  {author} {\bibinfo {author} {\bibfnamefont {X.}~\bibnamefont
  {Wan}}, \bibinfo {author} {\bibfnamefont {A.~M.}\ \bibnamefont {Turner}},
  \bibinfo {author} {\bibfnamefont {A.}~\bibnamefont {Vishwanath}}, \ and\
  \bibinfo {author} {\bibfnamefont {S.~Y.}\ \bibnamefont {Savrasov}},\ }\href
  {\doibase 10.1103/PhysRevB.83.205101} {\bibfield  {journal} {\bibinfo
  {journal} {Phys. Rev. B}\ }\textbf {\bibinfo {volume} {83}},\ \bibinfo
  {pages} {205101} (\bibinfo {year} {2011})}\BibitemShut {NoStop}%
\bibitem [{\citenamefont {Burkov}\ and\ \citenamefont
  {Balents}(2011)}]{Burkov2011weyl}%
  \BibitemOpen
  \bibfield  {author} {\bibinfo {author} {\bibfnamefont {A.~A.}\ \bibnamefont
  {Burkov}}\ and\ \bibinfo {author} {\bibfnamefont {L.}~\bibnamefont
  {Balents}},\ }\href {\doibase 10.1103/PhysRevLett.107.127205} {\bibfield
  {journal} {\bibinfo  {journal} {Phys. Rev. Lett.}\ }\textbf {\bibinfo
  {volume} {107}},\ \bibinfo {pages} {127205} (\bibinfo {year}
  {2011})}\BibitemShut {NoStop}%
\bibitem [{\citenamefont {Weng}\ \emph {et~al.}(2015)\citenamefont {Weng},
  \citenamefont {Fang}, \citenamefont {Fang}, \citenamefont {Bernevig},\ and\
  \citenamefont {Dai}}]{Weng2015weyl}%
  \BibitemOpen
  \bibfield  {author} {\bibinfo {author} {\bibfnamefont {H.}~\bibnamefont
  {Weng}}, \bibinfo {author} {\bibfnamefont {C.}~\bibnamefont {Fang}}, \bibinfo
  {author} {\bibfnamefont {Z.}~\bibnamefont {Fang}}, \bibinfo {author}
  {\bibfnamefont {B.~A.}\ \bibnamefont {Bernevig}}, \ and\ \bibinfo {author}
  {\bibfnamefont {X.}~\bibnamefont {Dai}},\ }\href {\doibase
  10.1103/PhysRevX.5.011029} {\bibfield  {journal} {\bibinfo  {journal} {Phys.
  Rev. X}\ }\textbf {\bibinfo {volume} {5}},\ \bibinfo {pages} {011029}
  (\bibinfo {year} {2015})}\BibitemShut {NoStop}%
\bibitem [{\citenamefont {Lu}\ \emph {et~al.}(2015)\citenamefont {Lu},
  \citenamefont {Wang}, \citenamefont {Ye}, \citenamefont {Ran}, \citenamefont
  {Fu}, \citenamefont {Joannopoulos},\ and\ \citenamefont
  {Solja{\v{c}}i{\'c}}}]{Lu2015weyl}%
  \BibitemOpen
  \bibfield  {author} {\bibinfo {author} {\bibfnamefont {L.}~\bibnamefont
  {Lu}}, \bibinfo {author} {\bibfnamefont {Z.}~\bibnamefont {Wang}}, \bibinfo
  {author} {\bibfnamefont {D.}~\bibnamefont {Ye}}, \bibinfo {author}
  {\bibfnamefont {L.}~\bibnamefont {Ran}}, \bibinfo {author} {\bibfnamefont
  {L.}~\bibnamefont {Fu}}, \bibinfo {author} {\bibfnamefont {J.~D.}\
  \bibnamefont {Joannopoulos}}, \ and\ \bibinfo {author} {\bibfnamefont
  {M.}~\bibnamefont {Solja{\v{c}}i{\'c}}},\ }\href {\doibase
  10.1126/science.aaa9273} {\bibfield  {journal} {\bibinfo  {journal}
  {Science}\ }\textbf {\bibinfo {volume} {349}},\ \bibinfo {pages} {622}
  (\bibinfo {year} {2015})}\BibitemShut {NoStop}%
\bibitem [{\citenamefont {Xu}\ \emph {et~al.}(2015{\natexlab{b}})\citenamefont
  {Xu}, \citenamefont {Belopolski}, \citenamefont {Alidoust}, \citenamefont
  {Neupane}, \citenamefont {Bian}, \citenamefont {Zhang}, \citenamefont
  {Sankar}, \citenamefont {Chang}, \citenamefont {Yuan}, \citenamefont {Lee}
  \emph {et~al.}}]{Xu2015TaAs}%
  \BibitemOpen
  \bibfield  {author} {\bibinfo {author} {\bibfnamefont {S.-Y.}\ \bibnamefont
  {Xu}}, \bibinfo {author} {\bibfnamefont {I.}~\bibnamefont {Belopolski}},
  \bibinfo {author} {\bibfnamefont {N.}~\bibnamefont {Alidoust}}, \bibinfo
  {author} {\bibfnamefont {M.}~\bibnamefont {Neupane}}, \bibinfo {author}
  {\bibfnamefont {G.}~\bibnamefont {Bian}}, \bibinfo {author} {\bibfnamefont
  {C.}~\bibnamefont {Zhang}}, \bibinfo {author} {\bibfnamefont
  {R.}~\bibnamefont {Sankar}}, \bibinfo {author} {\bibfnamefont
  {G.}~\bibnamefont {Chang}}, \bibinfo {author} {\bibfnamefont
  {Z.}~\bibnamefont {Yuan}}, \bibinfo {author} {\bibfnamefont {C.-C.}\
  \bibnamefont {Lee}},  \emph {et~al.},\ }\href {\doibase
  10.1126/science.aaa9297} {\bibfield  {journal} {\bibinfo  {journal}
  {Science}\ }\textbf {\bibinfo {volume} {349}},\ \bibinfo {pages} {613}
  (\bibinfo {year} {2015}{\natexlab{b}})}\BibitemShut {NoStop}%
\bibitem [{\citenamefont {Lv}\ \emph {et~al.}(2015)\citenamefont {Lv},
  \citenamefont {Weng}, \citenamefont {Fu}, \citenamefont {Wang}, \citenamefont
  {Miao}, \citenamefont {Ma}, \citenamefont {Richard}, \citenamefont {Huang},
  \citenamefont {Zhao}, \citenamefont {Chen}, \citenamefont {Fang},
  \citenamefont {Dai}, \citenamefont {Qian},\ and\ \citenamefont
  {Ding}}]{Lv2015TaAs}%
  \BibitemOpen
  \bibfield  {author} {\bibinfo {author} {\bibfnamefont {B.~Q.}\ \bibnamefont
  {Lv}}, \bibinfo {author} {\bibfnamefont {H.~M.}\ \bibnamefont {Weng}},
  \bibinfo {author} {\bibfnamefont {B.~B.}\ \bibnamefont {Fu}}, \bibinfo
  {author} {\bibfnamefont {X.~P.}\ \bibnamefont {Wang}}, \bibinfo {author}
  {\bibfnamefont {H.}~\bibnamefont {Miao}}, \bibinfo {author} {\bibfnamefont
  {J.}~\bibnamefont {Ma}}, \bibinfo {author} {\bibfnamefont {P.}~\bibnamefont
  {Richard}}, \bibinfo {author} {\bibfnamefont {X.~C.}\ \bibnamefont {Huang}},
  \bibinfo {author} {\bibfnamefont {L.~X.}\ \bibnamefont {Zhao}}, \bibinfo
  {author} {\bibfnamefont {G.~F.}\ \bibnamefont {Chen}}, \bibinfo {author}
  {\bibfnamefont {Z.}~\bibnamefont {Fang}}, \bibinfo {author} {\bibfnamefont
  {X.}~\bibnamefont {Dai}}, \bibinfo {author} {\bibfnamefont {T.}~\bibnamefont
  {Qian}}, \ and\ \bibinfo {author} {\bibfnamefont {H.}~\bibnamefont {Ding}},\
  }\href {\doibase 10.1103/PhysRevX.5.031013} {\bibfield  {journal} {\bibinfo
  {journal} {Phys. Rev. X}\ }\textbf {\bibinfo {volume} {5}},\ \bibinfo {pages}
  {031013} (\bibinfo {year} {2015})}\BibitemShut {NoStop}%
\bibitem [{\citenamefont {Yang}\ \emph {et~al.}(2015)\citenamefont {Yang},
  \citenamefont {Liu}, \citenamefont {Sun}, \citenamefont {Peng}, \citenamefont
  {Yang}, \citenamefont {Zhang}, \citenamefont {Zhou}, \citenamefont {Zhang},
  \citenamefont {Guo}, \citenamefont {Rahn} \emph {et~al.}}]{Yang2015weyl}%
  \BibitemOpen
  \bibfield  {author} {\bibinfo {author} {\bibfnamefont {L.}~\bibnamefont
  {Yang}}, \bibinfo {author} {\bibfnamefont {Z.}~\bibnamefont {Liu}}, \bibinfo
  {author} {\bibfnamefont {Y.}~\bibnamefont {Sun}}, \bibinfo {author}
  {\bibfnamefont {H.}~\bibnamefont {Peng}}, \bibinfo {author} {\bibfnamefont
  {H.}~\bibnamefont {Yang}}, \bibinfo {author} {\bibfnamefont {T.}~\bibnamefont
  {Zhang}}, \bibinfo {author} {\bibfnamefont {B.}~\bibnamefont {Zhou}},
  \bibinfo {author} {\bibfnamefont {Y.}~\bibnamefont {Zhang}}, \bibinfo
  {author} {\bibfnamefont {Y.}~\bibnamefont {Guo}}, \bibinfo {author}
  {\bibfnamefont {M.}~\bibnamefont {Rahn}},  \emph {et~al.},\ }\href {\doibase
  10.1038/nphys3425} {\bibfield  {journal} {\bibinfo  {journal} {Nature
  physics}\ }\textbf {\bibinfo {volume} {11}},\ \bibinfo {pages} {728}
  (\bibinfo {year} {2015})}\BibitemShut {NoStop}%
\bibitem [{\citenamefont {Huang}\ \emph {et~al.}(2015)\citenamefont {Huang},
  \citenamefont {Zhao}, \citenamefont {Long}, \citenamefont {Wang},
  \citenamefont {Chen}, \citenamefont {Yang}, \citenamefont {Liang},
  \citenamefont {Xue}, \citenamefont {Weng}, \citenamefont {Fang},
  \citenamefont {Dai},\ and\ \citenamefont {Chen}}]{Huang2015weyl}%
  \BibitemOpen
  \bibfield  {author} {\bibinfo {author} {\bibfnamefont {X.}~\bibnamefont
  {Huang}}, \bibinfo {author} {\bibfnamefont {L.}~\bibnamefont {Zhao}},
  \bibinfo {author} {\bibfnamefont {Y.}~\bibnamefont {Long}}, \bibinfo {author}
  {\bibfnamefont {P.}~\bibnamefont {Wang}}, \bibinfo {author} {\bibfnamefont
  {D.}~\bibnamefont {Chen}}, \bibinfo {author} {\bibfnamefont {Z.}~\bibnamefont
  {Yang}}, \bibinfo {author} {\bibfnamefont {H.}~\bibnamefont {Liang}},
  \bibinfo {author} {\bibfnamefont {M.}~\bibnamefont {Xue}}, \bibinfo {author}
  {\bibfnamefont {H.}~\bibnamefont {Weng}}, \bibinfo {author} {\bibfnamefont
  {Z.}~\bibnamefont {Fang}}, \bibinfo {author} {\bibfnamefont {X.}~\bibnamefont
  {Dai}}, \ and\ \bibinfo {author} {\bibfnamefont {G.}~\bibnamefont {Chen}},\
  }\href {\doibase 10.1103/PhysRevX.5.031023} {\bibfield  {journal} {\bibinfo
  {journal} {Phys. Rev. X}\ }\textbf {\bibinfo {volume} {5}},\ \bibinfo {pages}
  {031023} (\bibinfo {year} {2015})}\BibitemShut {NoStop}%
\bibitem [{\citenamefont {Di-Fei}\ \emph {et~al.}(2015)\citenamefont {Di-Fei},
  \citenamefont {Yong-Ping}, \citenamefont {Zhen}, \citenamefont {Yu-Peng},
  \citenamefont {Xiao-Hai}, \citenamefont {Qi}, \citenamefont {Pavel},
  \citenamefont {Zhu-An}, \citenamefont {Xian-Gang},\ and\ \citenamefont
  {Dong-Lai}}]{Xu2015weyl}%
  \BibitemOpen
  \bibfield  {author} {\bibinfo {author} {\bibfnamefont {X.}~\bibnamefont
  {Di-Fei}}, \bibinfo {author} {\bibfnamefont {D.}~\bibnamefont {Yong-Ping}},
  \bibinfo {author} {\bibfnamefont {W.}~\bibnamefont {Zhen}}, \bibinfo {author}
  {\bibfnamefont {L.}~\bibnamefont {Yu-Peng}}, \bibinfo {author} {\bibfnamefont
  {N.}~\bibnamefont {Xiao-Hai}}, \bibinfo {author} {\bibfnamefont
  {Y.}~\bibnamefont {Qi}}, \bibinfo {author} {\bibfnamefont {D.}~\bibnamefont
  {Pavel}}, \bibinfo {author} {\bibfnamefont {X.}~\bibnamefont {Zhu-An}},
  \bibinfo {author} {\bibfnamefont {W.}~\bibnamefont {Xian-Gang}}, \ and\
  \bibinfo {author} {\bibfnamefont {F.}~\bibnamefont {Dong-Lai}},\ }\href
  {http://stacks.iop.org/0256-307X/32/i=10/a=107101} {\bibfield  {journal}
  {\bibinfo  {journal} {Chinese Physics Letters}\ }\textbf {\bibinfo {volume}
  {32}},\ \bibinfo {pages} {107101} (\bibinfo {year} {2015})}\BibitemShut
  {NoStop}%
\bibitem [{\citenamefont {{Zeng}}\ \emph {et~al.}(2015)\citenamefont {{Zeng}},
  \citenamefont {{Fang}}, \citenamefont {{Chang}}, \citenamefont {{Chen}},
  \citenamefont {{Hsieh}}, \citenamefont {{Bansil}}, \citenamefont {{Lin}},\
  and\ \citenamefont {{Fu}}}]{Fuliang2015}%
  \BibitemOpen
  \bibfield  {author} {\bibinfo {author} {\bibfnamefont {M.}~\bibnamefont
  {{Zeng}}}, \bibinfo {author} {\bibfnamefont {C.}~\bibnamefont {{Fang}}},
  \bibinfo {author} {\bibfnamefont {G.}~\bibnamefont {{Chang}}}, \bibinfo
  {author} {\bibfnamefont {Y.-A.}\ \bibnamefont {{Chen}}}, \bibinfo {author}
  {\bibfnamefont {T.}~\bibnamefont {{Hsieh}}}, \bibinfo {author} {\bibfnamefont
  {A.}~\bibnamefont {{Bansil}}}, \bibinfo {author} {\bibfnamefont
  {H.}~\bibnamefont {{Lin}}}, \ and\ \bibinfo {author} {\bibfnamefont
  {L.}~\bibnamefont {{Fu}}},\ }\href@noop {} {\  (\bibinfo {year} {2015})},\
  \Eprint {http://arxiv.org/abs/1504.03492} {arXiv:1504.03492} \BibitemShut
  {NoStop}%
\bibitem [{\citenamefont {Kasuya}\ \emph {et~al.}(1993)\citenamefont {Kasuya},
  \citenamefont {Sera},\ and\ \citenamefont {Suzuki}}]{Kasuya1993}%
  \BibitemOpen
  \bibfield  {author} {\bibinfo {author} {\bibfnamefont {T.}~\bibnamefont
  {Kasuya}}, \bibinfo {author} {\bibfnamefont {M.}~\bibnamefont {Sera}}, \ and\
  \bibinfo {author} {\bibfnamefont {T.}~\bibnamefont {Suzuki}},\ }\href
  {\doibase 10.1143/JPSJ.62.2561} {\bibfield  {journal} {\bibinfo  {journal}
  {J. Phys. Soc. Japan}\ }\textbf {\bibinfo {volume} {62}},\ \bibinfo {pages}
  {2561} (\bibinfo {year} {1993})}\BibitemShut {NoStop}%
\bibitem [{\citenamefont {Tafti}\ \emph {et~al.}(2016)\citenamefont {Tafti},
  \citenamefont {Gibson}, \citenamefont {Kushwaha}, \citenamefont
  {Haldolaarachchige},\ and\ \citenamefont {Cava}}]{Tafti2016}%
  \BibitemOpen
  \bibfield  {author} {\bibinfo {author} {\bibfnamefont {F.~F.}\ \bibnamefont
  {Tafti}}, \bibinfo {author} {\bibfnamefont {Q.~D.}\ \bibnamefont {Gibson}},
  \bibinfo {author} {\bibfnamefont {S.~K.}\ \bibnamefont {Kushwaha}}, \bibinfo
  {author} {\bibfnamefont {N.}~\bibnamefont {Haldolaarachchige}}, \ and\
  \bibinfo {author} {\bibfnamefont {R.~J.}\ \bibnamefont {Cava}},\ }\href
  {\doibase 10.1038/nphys3581} {\bibfield  {journal} {\bibinfo  {journal}
  {Nat.\ Phys.}\ }\textbf {\bibinfo {volume} {12}},\ \bibinfo {pages} {272}
  (\bibinfo {year} {2016})}\BibitemShut {NoStop}%
\bibitem [{\citenamefont {Zeng}\ \emph {et~al.}(2016)\citenamefont {Zeng},
  \citenamefont {Lou}, \citenamefont {Wu}, \citenamefont {Guo}, \citenamefont
  {Kong}, \citenamefont {Zhong}, \citenamefont {Ma}, \citenamefont {Fu},
  \citenamefont {Richard}, \citenamefont {Wang}, \citenamefont {Liu},
  \citenamefont {Lu}, \citenamefont {Sun}, \citenamefont {Wang}, \citenamefont
  {Wang}, \citenamefont {Shi}, \citenamefont {Lei}, \citenamefont {Liu},
  \citenamefont {Wang}, \citenamefont {Qian}, \citenamefont {Luo},\ and\
  \citenamefont {Ding}}]{Zeng2016}%
  \BibitemOpen
  \bibfield  {author} {\bibinfo {author} {\bibfnamefont {L.-K.}\ \bibnamefont
  {Zeng}}, \bibinfo {author} {\bibfnamefont {R.}~\bibnamefont {Lou}}, \bibinfo
  {author} {\bibfnamefont {D.-S.}\ \bibnamefont {Wu}}, \bibinfo {author}
  {\bibfnamefont {P.-J.}\ \bibnamefont {Guo}}, \bibinfo {author} {\bibfnamefont
  {L.-Y.}\ \bibnamefont {Kong}}, \bibinfo {author} {\bibfnamefont {Y.-G.}\
  \bibnamefont {Zhong}}, \bibinfo {author} {\bibfnamefont {J.-Z.}\ \bibnamefont
  {Ma}}, \bibinfo {author} {\bibfnamefont {B.-B.}\ \bibnamefont {Fu}}, \bibinfo
  {author} {\bibfnamefont {P.}~\bibnamefont {Richard}}, \bibinfo {author}
  {\bibfnamefont {P.}~\bibnamefont {Wang}}, \bibinfo {author} {\bibfnamefont
  {G.~T.}\ \bibnamefont {Liu}}, \bibinfo {author} {\bibfnamefont
  {L.}~\bibnamefont {Lu}}, \bibinfo {author} {\bibfnamefont {S.-S.}\
  \bibnamefont {Sun}}, \bibinfo {author} {\bibfnamefont {Q.}~\bibnamefont
  {Wang}}, \bibinfo {author} {\bibfnamefont {L.}~\bibnamefont {Wang}}, \bibinfo
  {author} {\bibfnamefont {Y.-G.}\ \bibnamefont {Shi}}, \bibinfo {author}
  {\bibfnamefont {H.-C.}\ \bibnamefont {Lei}}, \bibinfo {author} {\bibfnamefont
  {K.}~\bibnamefont {Liu}}, \bibinfo {author} {\bibfnamefont {S.-C.}\
  \bibnamefont {Wang}}, \bibinfo {author} {\bibfnamefont {T.}~\bibnamefont
  {Qian}}, \bibinfo {author} {\bibfnamefont {J.-L.}\ \bibnamefont {Luo}}, \
  and\ \bibinfo {author} {\bibfnamefont {H.}~\bibnamefont {Ding}},\ }\href@noop
  {} {\  (\bibinfo {year} {2016})},\ \Eprint {http://arxiv.org/abs/1604.08142}
  {arXiv:1604.08142} \BibitemShut {NoStop}%
\bibitem [{\citenamefont {Stepanov}\ \emph {et~al.}(2015)\citenamefont
  {Stepanov}, \citenamefont {Morozova}, \citenamefont {Kar'kin}, \citenamefont
  {Golubkov},\ and\ \citenamefont {Kaminskii}}]{Stepanov2015}%
  \BibitemOpen
  \bibfield  {author} {\bibinfo {author} {\bibfnamefont {N.~N.}\ \bibnamefont
  {Stepanov}}, \bibinfo {author} {\bibfnamefont {N.~V.}\ \bibnamefont
  {Morozova}}, \bibinfo {author} {\bibfnamefont {A.~E.}\ \bibnamefont
  {Kar'kin}}, \bibinfo {author} {\bibfnamefont {A.~V.}\ \bibnamefont
  {Golubkov}}, \ and\ \bibinfo {author} {\bibfnamefont {V.~V.}\ \bibnamefont
  {Kaminskii}},\ }\href {\doibase 10.1134/S106378341512032X} {\bibfield
  {journal} {\bibinfo  {journal} {Physics of the Solid State}\ }\textbf
  {\bibinfo {volume} {57}},\ \bibinfo {pages} {2369} (\bibinfo {year}
  {2015})}\BibitemShut {NoStop}%
\bibitem [{\citenamefont {Sun}\ \emph {et~al.}(2016)\citenamefont {Sun},
  \citenamefont {Wang}, \citenamefont {Guo}, \citenamefont {Liu},\ and\
  \citenamefont {Lei}}]{Sun2016}%
  \BibitemOpen
  \bibfield  {author} {\bibinfo {author} {\bibfnamefont {S.}~\bibnamefont
  {Sun}}, \bibinfo {author} {\bibfnamefont {Q.}~\bibnamefont {Wang}}, \bibinfo
  {author} {\bibfnamefont {P.-J.}\ \bibnamefont {Guo}}, \bibinfo {author}
  {\bibfnamefont {K.}~\bibnamefont {Liu}}, \ and\ \bibinfo {author}
  {\bibfnamefont {H.}~\bibnamefont {Lei}},\ }\href
  {http://stacks.iop.org/1367-2630/18/i=8/a=082002} {\bibfield  {journal}
  {\bibinfo  {journal} {New Journal of Physics}\ }\textbf {\bibinfo {volume}
  {18}},\ \bibinfo {pages} {082002} (\bibinfo {year} {2016})}\BibitemShut
  {NoStop}%
\bibitem [{\citenamefont {Wu}\ \emph {et~al.}(2016)\citenamefont {Wu},
  \citenamefont {Kong}, \citenamefont {Wang}, \citenamefont {Johnson},
  \citenamefont {Mou}, \citenamefont {Huang}, \citenamefont {Schrunk},
  \citenamefont {Bud'ko}, \citenamefont {Canfield},\ and\ \citenamefont
  {Kaminski}}]{Wu2016}%
  \BibitemOpen
  \bibfield  {author} {\bibinfo {author} {\bibfnamefont {Y.}~\bibnamefont
  {Wu}}, \bibinfo {author} {\bibfnamefont {T.}~\bibnamefont {Kong}}, \bibinfo
  {author} {\bibfnamefont {L.-L.}\ \bibnamefont {Wang}}, \bibinfo {author}
  {\bibfnamefont {D.~D.}\ \bibnamefont {Johnson}}, \bibinfo {author}
  {\bibfnamefont {D.}~\bibnamefont {Mou}}, \bibinfo {author} {\bibfnamefont
  {L.}~\bibnamefont {Huang}}, \bibinfo {author} {\bibfnamefont
  {B.}~\bibnamefont {Schrunk}}, \bibinfo {author} {\bibfnamefont {S.~L.}\
  \bibnamefont {Bud'ko}}, \bibinfo {author} {\bibfnamefont {P.~C.}\
  \bibnamefont {Canfield}}, \ and\ \bibinfo {author} {\bibfnamefont
  {A.}~\bibnamefont {Kaminski}},\ }\href@noop {} {\  (\bibinfo {year}
  {2016})},\ \Eprint {http://arxiv.org/abs/1604.08945} {arXiv:1604.08945}
  \BibitemShut {NoStop}%
\bibitem [{\citenamefont {Kumar}\ \emph {et~al.}(2016)\citenamefont {Kumar},
  \citenamefont {Shekhar}, \citenamefont {Wu}, \citenamefont {Leermakers},
  \citenamefont {Young}, \citenamefont {Zeitler}, \citenamefont {Yan},\ and\
  \citenamefont {Felser}}]{Kumar2016}%
  \BibitemOpen
  \bibfield  {author} {\bibinfo {author} {\bibfnamefont {N.}~\bibnamefont
  {Kumar}}, \bibinfo {author} {\bibfnamefont {C.}~\bibnamefont {Shekhar}},
  \bibinfo {author} {\bibfnamefont {S.-C.}\ \bibnamefont {Wu}}, \bibinfo
  {author} {\bibfnamefont {I.}~\bibnamefont {Leermakers}}, \bibinfo {author}
  {\bibfnamefont {O.}~\bibnamefont {Young}}, \bibinfo {author} {\bibfnamefont
  {U.}~\bibnamefont {Zeitler}}, \bibinfo {author} {\bibfnamefont
  {B.}~\bibnamefont {Yan}}, \ and\ \bibinfo {author} {\bibfnamefont
  {C.}~\bibnamefont {Felser}},\ }\href {\doibase 10.1103/PhysRevB.93.241106}
  {\bibfield  {journal} {\bibinfo  {journal} {Phys. Rev. B}\ }\textbf {\bibinfo
  {volume} {93}},\ \bibinfo {pages} {241106} (\bibinfo {year}
  {2016})}\BibitemShut {NoStop}%
\bibitem [{\citenamefont {Ghimire}\ \emph {et~al.}(2016)\citenamefont
  {Ghimire}, \citenamefont {Botana}, \citenamefont {Phelan}, \citenamefont
  {Zheng},\ and\ \citenamefont {Mitchell}}]{Ghimire2016}%
  \BibitemOpen
  \bibfield  {author} {\bibinfo {author} {\bibfnamefont {N.~J.}\ \bibnamefont
  {Ghimire}}, \bibinfo {author} {\bibfnamefont {A.~S.}\ \bibnamefont {Botana}},
  \bibinfo {author} {\bibfnamefont {D.}~\bibnamefont {Phelan}}, \bibinfo
  {author} {\bibfnamefont {H.}~\bibnamefont {Zheng}}, \ and\ \bibinfo {author}
  {\bibfnamefont {J.~F.}\ \bibnamefont {Mitchell}},\ }\href {\doibase
  10.1088/0953-8984/28/23/235601} {\bibfield  {journal} {\bibinfo  {journal}
  {J. Phys.: Condens. Matt.}\ }\textbf {\bibinfo {volume} {28}},\ \bibinfo
  {pages} {235601} (\bibinfo {year} {2016})},\ \Eprint
  {http://arxiv.org/abs/1604.04232} {arXiv:1604.04232} \BibitemShut {NoStop}%
\bibitem [{\citenamefont {Yu}\ \emph {et~al.}(2016)\citenamefont {Yu},
  \citenamefont {Wang}, \citenamefont {Xu},\ and\ \citenamefont
  {Xia}}]{Yu2016}%
  \BibitemOpen
  \bibfield  {author} {\bibinfo {author} {\bibfnamefont {Q.-H.}\ \bibnamefont
  {Yu}}, \bibinfo {author} {\bibfnamefont {Y.-Y.}\ \bibnamefont {Wang}},
  \bibinfo {author} {\bibfnamefont {S.}~\bibnamefont {Xu}}, \ and\ \bibinfo
  {author} {\bibfnamefont {T.-L.}\ \bibnamefont {Xia}},\ }\href@noop {} {\
  (\bibinfo {year} {2016})},\ \Eprint {http://arxiv.org/abs/1604.05912}
  {arXiv:1604.05912} \BibitemShut {NoStop}%
\bibitem [{\citenamefont {Orest~Pavlosiuk}(2016)}]{Pavlosiuk2016}%
  \BibitemOpen
  \bibfield  {author} {\bibinfo {author} {\bibfnamefont {P.~W.}\ \bibnamefont
  {Orest~Pavlosiuk}, \bibfnamefont {Przemys{\l}aw~Swatek}},\ }\href@noop {} {\
  (\bibinfo {year} {2016})},\ \Eprint {http://arxiv.org/abs/1604.06945}
  {arXiv:1604.06945} \BibitemShut {NoStop}%
\bibitem [{\citenamefont {Abrikosov}(1998)}]{Abrikosov1998}%
  \BibitemOpen
  \bibfield  {author} {\bibinfo {author} {\bibfnamefont {A.~A.}\ \bibnamefont
  {Abrikosov}},\ }\href {\doibase 10.1103/PhysRevB.58.2788} {\bibfield
  {journal} {\bibinfo  {journal} {Phys. Rev. B}\ }\textbf {\bibinfo {volume}
  {58}},\ \bibinfo {pages} {2788} (\bibinfo {year} {1998})}\BibitemShut
  {NoStop}%
\bibitem [{\citenamefont {Yang}\ \emph {et~al.}(1999)\citenamefont {Yang},
  \citenamefont {Liu}, \citenamefont {Hong}, \citenamefont {Reich},
  \citenamefont {Searson},\ and\ \citenamefont {Chien}}]{Yang1999}%
  \BibitemOpen
  \bibfield  {author} {\bibinfo {author} {\bibfnamefont {F.~Y.}\ \bibnamefont
  {Yang}}, \bibinfo {author} {\bibfnamefont {K.}~\bibnamefont {Liu}}, \bibinfo
  {author} {\bibfnamefont {K.}~\bibnamefont {Hong}}, \bibinfo {author}
  {\bibfnamefont {D.~H.}\ \bibnamefont {Reich}}, \bibinfo {author}
  {\bibfnamefont {P.~C.}\ \bibnamefont {Searson}}, \ and\ \bibinfo {author}
  {\bibfnamefont {C.~L.}\ \bibnamefont {Chien}},\ }\href {\doibase
  10.1126/science.284.5418.1335} {\bibfield  {journal} {\bibinfo  {journal}
  {Science}\ }\textbf {\bibinfo {volume} {284}},\ \bibinfo {pages} {1335}
  (\bibinfo {year} {1999})}\BibitemShut {NoStop}%
\bibitem [{\citenamefont {Mun}\ \emph {et~al.}(2012)\citenamefont {Mun},
  \citenamefont {Ko}, \citenamefont {Miller}, \citenamefont {Samolyuk},
  \citenamefont {Bud'ko},\ and\ \citenamefont {Canfield}}]{Mun2012}%
  \BibitemOpen
  \bibfield  {author} {\bibinfo {author} {\bibfnamefont {E.}~\bibnamefont
  {Mun}}, \bibinfo {author} {\bibfnamefont {H.}~\bibnamefont {Ko}}, \bibinfo
  {author} {\bibfnamefont {G.~J.}\ \bibnamefont {Miller}}, \bibinfo {author}
  {\bibfnamefont {G.~D.}\ \bibnamefont {Samolyuk}}, \bibinfo {author}
  {\bibfnamefont {S.~L.}\ \bibnamefont {Bud'ko}}, \ and\ \bibinfo {author}
  {\bibfnamefont {P.~C.}\ \bibnamefont {Canfield}},\ }\href {\doibase
  10.1103/PhysRevB.85.035135} {\bibfield  {journal} {\bibinfo  {journal} {Phys.
  Rev. B}\ }\textbf {\bibinfo {volume} {85}},\ \bibinfo {pages} {035135}
  (\bibinfo {year} {2012})}\BibitemShut {NoStop}%
\bibitem [{\citenamefont {Jiang}\ \emph {et~al.}(2015)\citenamefont {Jiang},
  \citenamefont {Tang}, \citenamefont {Pan}, \citenamefont {Liu}, \citenamefont
  {Niu}, \citenamefont {Wang}, \citenamefont {Xu}, \citenamefont {Yang},
  \citenamefont {Xie}, \citenamefont {Song}, \citenamefont {Dudin},
  \citenamefont {Kim}, \citenamefont {Hoesch}, \citenamefont {Das},
  \citenamefont {Vobornik}, \citenamefont {Wan},\ and\ \citenamefont
  {Feng}}]{Jiang2015}%
  \BibitemOpen
  \bibfield  {author} {\bibinfo {author} {\bibfnamefont {J.}~\bibnamefont
  {Jiang}}, \bibinfo {author} {\bibfnamefont {F.}~\bibnamefont {Tang}},
  \bibinfo {author} {\bibfnamefont {X.~C.}\ \bibnamefont {Pan}}, \bibinfo
  {author} {\bibfnamefont {H.~M.}\ \bibnamefont {Liu}}, \bibinfo {author}
  {\bibfnamefont {X.~H.}\ \bibnamefont {Niu}}, \bibinfo {author} {\bibfnamefont
  {Y.~X.}\ \bibnamefont {Wang}}, \bibinfo {author} {\bibfnamefont {D.~F.}\
  \bibnamefont {Xu}}, \bibinfo {author} {\bibfnamefont {H.~F.}\ \bibnamefont
  {Yang}}, \bibinfo {author} {\bibfnamefont {B.~P.}\ \bibnamefont {Xie}},
  \bibinfo {author} {\bibfnamefont {F.~Q.}\ \bibnamefont {Song}}, \bibinfo
  {author} {\bibfnamefont {P.}~\bibnamefont {Dudin}}, \bibinfo {author}
  {\bibfnamefont {T.~K.}\ \bibnamefont {Kim}}, \bibinfo {author} {\bibfnamefont
  {M.}~\bibnamefont {Hoesch}}, \bibinfo {author} {\bibfnamefont {P.~K.}\
  \bibnamefont {Das}}, \bibinfo {author} {\bibfnamefont {I.}~\bibnamefont
  {Vobornik}}, \bibinfo {author} {\bibfnamefont {X.~G.}\ \bibnamefont {Wan}}, \
  and\ \bibinfo {author} {\bibfnamefont {D.~L.}\ \bibnamefont {Feng}},\ }\href
  {\doibase 10.1103/PhysRevLett.115.166601} {\bibfield  {journal} {\bibinfo
  {journal} {Phys. Rev. Lett.}\ }\textbf {\bibinfo {volume} {115}},\ \bibinfo
  {pages} {166601} (\bibinfo {year} {2015})}\BibitemShut {NoStop}%
\bibitem [{\citenamefont {Liang}\ \emph {et~al.}(2015)\citenamefont {Liang},
  \citenamefont {Gibson}, \citenamefont {Ali}, \citenamefont {Liu},
  \citenamefont {Cava},\ and\ \citenamefont {Ong}}]{liang2015CdAs}%
  \BibitemOpen
  \bibfield  {author} {\bibinfo {author} {\bibfnamefont {T.}~\bibnamefont
  {Liang}}, \bibinfo {author} {\bibfnamefont {Q.}~\bibnamefont {Gibson}},
  \bibinfo {author} {\bibfnamefont {M.~N.}\ \bibnamefont {Ali}}, \bibinfo
  {author} {\bibfnamefont {M.}~\bibnamefont {Liu}}, \bibinfo {author}
  {\bibfnamefont {R.}~\bibnamefont {Cava}}, \ and\ \bibinfo {author}
  {\bibfnamefont {N.}~\bibnamefont {Ong}},\ }\href {\doibase 10.1038/nmat4143}
  {\bibfield  {journal} {\bibinfo  {journal} {Nature materials}\ }\textbf
  {\bibinfo {volume} {14}},\ \bibinfo {pages} {280} (\bibinfo {year}
  {2015})}\BibitemShut {NoStop}%
\bibitem [{\citenamefont {Guo}\ \emph {et~al.}(2016)\citenamefont {Guo},
  \citenamefont {Yang}, \citenamefont {Zhang}, \citenamefont {Liu},\ and\
  \citenamefont {Lu}}]{Guo2016}%
  \BibitemOpen
  \bibfield  {author} {\bibinfo {author} {\bibfnamefont {P.-J.}\ \bibnamefont
  {Guo}}, \bibinfo {author} {\bibfnamefont {H.-C.}\ \bibnamefont {Yang}},
  \bibinfo {author} {\bibfnamefont {B.-J.}\ \bibnamefont {Zhang}}, \bibinfo
  {author} {\bibfnamefont {K.}~\bibnamefont {Liu}}, \ and\ \bibinfo {author}
  {\bibfnamefont {Z.-Y.}\ \bibnamefont {Lu}},\ }\href {\doibase
  10.1103/PhysRevB.93.235142} {\bibfield  {journal} {\bibinfo  {journal} {Phys.
  Rev. B}\ }\textbf {\bibinfo {volume} {93}},\ \bibinfo {pages} {235142}
  (\bibinfo {year} {2016})}\BibitemShut {NoStop}%
\bibitem [{\citenamefont {Nayak}\ \emph {et~al.}(2016)\citenamefont {Nayak},
  \citenamefont {Wu}, \citenamefont {Kumar}, \citenamefont {Shekhar},
  \citenamefont {Singh}, \citenamefont {Fink}, \citenamefont {Rienks},
  \citenamefont {Fecher}, \citenamefont {Parkin}, \citenamefont {Yan},\ and\
  \citenamefont {Felser}}]{Nayak2016}%
  \BibitemOpen
  \bibfield  {author} {\bibinfo {author} {\bibfnamefont {J.}~\bibnamefont
  {Nayak}}, \bibinfo {author} {\bibfnamefont {S.-C.}\ \bibnamefont {Wu}},
  \bibinfo {author} {\bibfnamefont {N.}~\bibnamefont {Kumar}}, \bibinfo
  {author} {\bibfnamefont {C.}~\bibnamefont {Shekhar}}, \bibinfo {author}
  {\bibfnamefont {S.}~\bibnamefont {Singh}}, \bibinfo {author} {\bibfnamefont
  {J.}~\bibnamefont {Fink}}, \bibinfo {author} {\bibfnamefont {E.~E.~D.}\
  \bibnamefont {Rienks}}, \bibinfo {author} {\bibfnamefont {G.~H.}\
  \bibnamefont {Fecher}}, \bibinfo {author} {\bibfnamefont {S.~S.~P.}\
  \bibnamefont {Parkin}}, \bibinfo {author} {\bibfnamefont {B.}~\bibnamefont
  {Yan}}, \ and\ \bibinfo {author} {\bibfnamefont {C.}~\bibnamefont {Felser}},\
  }\href@noop {} {\  (\bibinfo {year} {2016})},\ \Eprint
  {http://arxiv.org/abs/1605.06997} {arXiv:1605.06997} \BibitemShut {NoStop}%
\bibitem [{\citenamefont {Canfield}\ and\ \citenamefont
  {Fisk}(1991)}]{Canfield1991}%
  \BibitemOpen
  \bibfield  {author} {\bibinfo {author} {\bibfnamefont {P.~C.}\ \bibnamefont
  {Canfield}}\ and\ \bibinfo {author} {\bibfnamefont {Z.}~\bibnamefont
  {Fisk}},\ }\href {\doibase 10.1080/13642819208215073} {\bibfield  {journal}
  {\bibinfo  {journal} {Phil.\ Mag.\ B}\ }\textbf {\bibinfo {volume} {65}},\
  \bibinfo {pages} {1117} (\bibinfo {year} {1991})}\BibitemShut {NoStop}%
\bibitem [{\citenamefont {Yoshida}\ \emph {et~al.}(2001)\citenamefont
  {Yoshida}, \citenamefont {Koyama}, \citenamefont {Tomimatsu}, \citenamefont
  {Shirakawa}, \citenamefont {Ochiai},\ and\ \citenamefont
  {Motokawa}}]{Yoshida2001}%
  \BibitemOpen
  \bibfield  {author} {\bibinfo {author} {\bibfnamefont {M.}~\bibnamefont
  {Yoshida}}, \bibinfo {author} {\bibfnamefont {K.}~\bibnamefont {Koyama}},
  \bibinfo {author} {\bibfnamefont {T.}~\bibnamefont {Tomimatsu}}, \bibinfo
  {author} {\bibfnamefont {M.}~\bibnamefont {Shirakawa}}, \bibinfo {author}
  {\bibfnamefont {A.}~\bibnamefont {Ochiai}}, \ and\ \bibinfo {author}
  {\bibfnamefont {M.}~\bibnamefont {Motokawa}},\ }\href {\doibase
  10.1143/JPSJ.70.2078} {\bibfield  {journal} {\bibinfo  {journal} {J. Phys.
  Soc. Japan}\ }\textbf {\bibinfo {volume} {70}},\ \bibinfo {pages} {2078}
  (\bibinfo {year} {2001})}\BibitemShut {NoStop}%
\bibitem [{\citenamefont {Kitazawa}\ \emph {et~al.}(1983)\citenamefont
  {Kitazawa}, \citenamefont {Suzuki}, \citenamefont {Sera}, \citenamefont
  {Oguro}, \citenamefont {Yanase}, \citenamefont {Hasegawa},\ and\
  \citenamefont {Kasuya}}]{Kitazawa1983}%
  \BibitemOpen
  \bibfield  {author} {\bibinfo {author} {\bibfnamefont {H.}~\bibnamefont
  {Kitazawa}}, \bibinfo {author} {\bibfnamefont {T.}~\bibnamefont {Suzuki}},
  \bibinfo {author} {\bibfnamefont {M.}~\bibnamefont {Sera}}, \bibinfo {author}
  {\bibfnamefont {I.}~\bibnamefont {Oguro}}, \bibinfo {author} {\bibfnamefont
  {A.}~\bibnamefont {Yanase}}, \bibinfo {author} {\bibfnamefont
  {A.}~\bibnamefont {Hasegawa}}, \ and\ \bibinfo {author} {\bibfnamefont
  {T.}~\bibnamefont {Kasuya}},\ }\href {\doibase 10.1016/0304-8853(83)90304-9}
  {\bibfield  {journal} {\bibinfo  {journal} {J.\ Magn.\ Magn.\ Mater.}\
  }\textbf {\bibinfo {volume} {31--34}},\ \bibinfo {pages} {421} (\bibinfo
  {year} {1983})}\BibitemShut {NoStop}%
\bibitem [{\citenamefont {Hasegawa}(1985)}]{Hasegawa1985}%
  \BibitemOpen
  \bibfield  {author} {\bibinfo {author} {\bibfnamefont {A.}~\bibnamefont
  {Hasegawa}},\ }\href {\doibase 10.1143/JPSJ.54.677} {\bibfield  {journal}
  {\bibinfo  {journal} {J. Phys. Soc. Japan}\ }\textbf {\bibinfo {volume}
  {54}},\ \bibinfo {pages} {677} (\bibinfo {year} {1985})}\BibitemShut
  {NoStop}%
\bibitem [{\citenamefont {Settai}\ \emph {et~al.}(1993)\citenamefont {Settai},
  \citenamefont {Goto}, \citenamefont {Sakatsume}, \citenamefont {Kwon},
  \citenamefont {Suzuki},\ and\ \citenamefont {Kasuya}}]{Settai1993}%
  \BibitemOpen
  \bibfield  {author} {\bibinfo {author} {\bibfnamefont {R.}~\bibnamefont
  {Settai}}, \bibinfo {author} {\bibfnamefont {T.}~\bibnamefont {Goto}},
  \bibinfo {author} {\bibfnamefont {S.}~\bibnamefont {Sakatsume}}, \bibinfo
  {author} {\bibfnamefont {Y.}~\bibnamefont {Kwon}}, \bibinfo {author}
  {\bibfnamefont {T.}~\bibnamefont {Suzuki}}, \ and\ \bibinfo {author}
  {\bibfnamefont {T.}~\bibnamefont {Kasuya}},\ }\href {\doibase
  10.1016/0921-4526(93)90527-D} {\bibfield  {journal} {\bibinfo  {journal}
  {Physica B: Condensed Matter}\ }\textbf {\bibinfo {volume} {186}},\ \bibinfo
  {pages} {176} (\bibinfo {year} {1993})}\BibitemShut {NoStop}%
\bibitem [{\citenamefont {Yoshida}\ \emph {et~al.}(2000)\citenamefont
  {Yoshida}, \citenamefont {Koyama}, \citenamefont {Sakon}, \citenamefont
  {Ochiai},\ and\ \citenamefont {Motokawa}}]{Yoshida2000}%
  \BibitemOpen
  \bibfield  {author} {\bibinfo {author} {\bibfnamefont {M.}~\bibnamefont
  {Yoshida}}, \bibinfo {author} {\bibfnamefont {K.}~\bibnamefont {Koyama}},
  \bibinfo {author} {\bibfnamefont {T.}~\bibnamefont {Sakon}}, \bibinfo
  {author} {\bibfnamefont {A.}~\bibnamefont {Ochiai}}, \ and\ \bibinfo {author}
  {\bibfnamefont {M.}~\bibnamefont {Motokawa}},\ }\href {\doibase
  10.1143/JPSJ.69.3629} {\bibfield  {journal} {\bibinfo  {journal} {J. Phys.
  Soc. Japan}\ }\textbf {\bibinfo {volume} {69}},\ \bibinfo {pages} {3629}
  (\bibinfo {year} {2000})}\BibitemShut {NoStop}%
\bibitem [{\citenamefont {Song}\ \emph {et~al.}(2016)\citenamefont {Song},
  \citenamefont {Yan}, \citenamefont {Ye}, \citenamefont {Ren}, \citenamefont
  {Xu}, \citenamefont {Tan}, \citenamefont {Niu}, \citenamefont {Xie},
  \citenamefont {Zhang}, \citenamefont {Peng}, \citenamefont {Xu},
  \citenamefont {Jiang},\ and\ \citenamefont {Feng}}]{Song2016}%
  \BibitemOpen
  \bibfield  {author} {\bibinfo {author} {\bibfnamefont {Q.}~\bibnamefont
  {Song}}, \bibinfo {author} {\bibfnamefont {Y.~J.}\ \bibnamefont {Yan}},
  \bibinfo {author} {\bibfnamefont {Z.~R.}\ \bibnamefont {Ye}}, \bibinfo
  {author} {\bibfnamefont {M.~Q.}\ \bibnamefont {Ren}}, \bibinfo {author}
  {\bibfnamefont {D.~F.}\ \bibnamefont {Xu}}, \bibinfo {author} {\bibfnamefont
  {S.~Y.}\ \bibnamefont {Tan}}, \bibinfo {author} {\bibfnamefont {X.~H.}\
  \bibnamefont {Niu}}, \bibinfo {author} {\bibfnamefont {B.~P.}\ \bibnamefont
  {Xie}}, \bibinfo {author} {\bibfnamefont {T.}~\bibnamefont {Zhang}}, \bibinfo
  {author} {\bibfnamefont {R.}~\bibnamefont {Peng}}, \bibinfo {author}
  {\bibfnamefont {H.~C.}\ \bibnamefont {Xu}}, \bibinfo {author} {\bibfnamefont
  {J.}~\bibnamefont {Jiang}}, \ and\ \bibinfo {author} {\bibfnamefont {D.~L.}\
  \bibnamefont {Feng}},\ }\href {\doibase 10.1103/PhysRevB.93.024508}
  {\bibfield  {journal} {\bibinfo  {journal} {Phys. Rev. B}\ }\textbf {\bibinfo
  {volume} {93}},\ \bibinfo {pages} {024508} (\bibinfo {year}
  {2016})}\BibitemShut {NoStop}%
\bibitem [{\citenamefont {Xu}\ \emph {et~al.}(2016)\citenamefont {Xu},
  \citenamefont {Shen}, \citenamefont {Zhu}, \citenamefont {Jiang},
  \citenamefont {Xie}, \citenamefont {Wang}, \citenamefont {Pan}, \citenamefont
  {Dudin}, \citenamefont {Kim}, \citenamefont {Hoesch}, \citenamefont {Zhao},
  \citenamefont {Wan},\ and\ \citenamefont {Feng}}]{Xu2016}%
  \BibitemOpen
  \bibfield  {author} {\bibinfo {author} {\bibfnamefont {D.~F.}\ \bibnamefont
  {Xu}}, \bibinfo {author} {\bibfnamefont {D.~W.}\ \bibnamefont {Shen}},
  \bibinfo {author} {\bibfnamefont {D.}~\bibnamefont {Zhu}}, \bibinfo {author}
  {\bibfnamefont {J.}~\bibnamefont {Jiang}}, \bibinfo {author} {\bibfnamefont
  {B.~P.}\ \bibnamefont {Xie}}, \bibinfo {author} {\bibfnamefont {Q.~S.}\
  \bibnamefont {Wang}}, \bibinfo {author} {\bibfnamefont {B.~Y.}\ \bibnamefont
  {Pan}}, \bibinfo {author} {\bibfnamefont {P.}~\bibnamefont {Dudin}}, \bibinfo
  {author} {\bibfnamefont {T.~K.}\ \bibnamefont {Kim}}, \bibinfo {author}
  {\bibfnamefont {M.}~\bibnamefont {Hoesch}}, \bibinfo {author} {\bibfnamefont
  {J.}~\bibnamefont {Zhao}}, \bibinfo {author} {\bibfnamefont {X.~G.}\
  \bibnamefont {Wan}}, \ and\ \bibinfo {author} {\bibfnamefont {D.~L.}\
  \bibnamefont {Feng}},\ }\href {\doibase 10.1103/PhysRevB.93.024506}
  {\bibfield  {journal} {\bibinfo  {journal} {Phys. Rev. B}\ }\textbf {\bibinfo
  {volume} {93}},\ \bibinfo {pages} {024506} (\bibinfo {year}
  {2016})}\BibitemShut {NoStop}%
\bibitem [{\citenamefont {Kumigashira}\ \emph {et~al.}(1998)\citenamefont
  {Kumigashira}, \citenamefont {Kim}, \citenamefont {Ito}, \citenamefont
  {Ashihara}, \citenamefont {Takahashi}, \citenamefont {Suzuki}, \citenamefont
  {Nishimura}, \citenamefont {Sakai}, \citenamefont {Kaneta},\ and\
  \citenamefont {Harima}}]{Kumigashira1998}%
  \BibitemOpen
  \bibfield  {author} {\bibinfo {author} {\bibfnamefont {H.}~\bibnamefont
  {Kumigashira}}, \bibinfo {author} {\bibfnamefont {H.-D.}\ \bibnamefont
  {Kim}}, \bibinfo {author} {\bibfnamefont {T.}~\bibnamefont {Ito}}, \bibinfo
  {author} {\bibfnamefont {A.}~\bibnamefont {Ashihara}}, \bibinfo {author}
  {\bibfnamefont {T.}~\bibnamefont {Takahashi}}, \bibinfo {author}
  {\bibfnamefont {T.}~\bibnamefont {Suzuki}}, \bibinfo {author} {\bibfnamefont
  {M.}~\bibnamefont {Nishimura}}, \bibinfo {author} {\bibfnamefont
  {O.}~\bibnamefont {Sakai}}, \bibinfo {author} {\bibfnamefont
  {Y.}~\bibnamefont {Kaneta}}, \ and\ \bibinfo {author} {\bibfnamefont
  {H.}~\bibnamefont {Harima}},\ }\href {\doibase 10.1103/PhysRevB.58.7675}
  {\bibfield  {journal} {\bibinfo  {journal} {Phys. Rev. B}\ }\textbf {\bibinfo
  {volume} {58}},\ \bibinfo {pages} {7675} (\bibinfo {year}
  {1998})}\BibitemShut {NoStop}%
\bibitem [{\citenamefont {Wang}\ and\ \citenamefont
  {Gedik}(2013)}]{Wang2013CD}%
  \BibitemOpen
  \bibfield  {author} {\bibinfo {author} {\bibfnamefont {Y.}~\bibnamefont
  {Wang}}\ and\ \bibinfo {author} {\bibfnamefont {N.}~\bibnamefont {Gedik}},\
  }\href {\doibase 10.1002/pssr.201206458} {\bibfield  {journal} {\bibinfo
  {journal} {physica status solidi (RRL) – Rapid Research Letters}\ }\textbf
  {\bibinfo {volume} {7}},\ \bibinfo {pages} {64} (\bibinfo {year}
  {2013})}\BibitemShut {NoStop}%
\bibitem [{\citenamefont {Park}\ \emph {et~al.}(2012)\citenamefont {Park},
  \citenamefont {Han}, \citenamefont {Kim}, \citenamefont {Koh}, \citenamefont
  {Kim}, \citenamefont {Lee}, \citenamefont {Choi}, \citenamefont {Han},
  \citenamefont {Lee}, \citenamefont {Hur}, \citenamefont {Arita},
  \citenamefont {Shimada}, \citenamefont {Namatame},\ and\ \citenamefont
  {Taniguchi}}]{Park2012CD}%
  \BibitemOpen
  \bibfield  {author} {\bibinfo {author} {\bibfnamefont {S.~R.}\ \bibnamefont
  {Park}}, \bibinfo {author} {\bibfnamefont {J.}~\bibnamefont {Han}}, \bibinfo
  {author} {\bibfnamefont {C.}~\bibnamefont {Kim}}, \bibinfo {author}
  {\bibfnamefont {Y.~Y.}\ \bibnamefont {Koh}}, \bibinfo {author} {\bibfnamefont
  {C.}~\bibnamefont {Kim}}, \bibinfo {author} {\bibfnamefont {H.}~\bibnamefont
  {Lee}}, \bibinfo {author} {\bibfnamefont {H.~J.}\ \bibnamefont {Choi}},
  \bibinfo {author} {\bibfnamefont {J.~H.}\ \bibnamefont {Han}}, \bibinfo
  {author} {\bibfnamefont {K.~D.}\ \bibnamefont {Lee}}, \bibinfo {author}
  {\bibfnamefont {N.~J.}\ \bibnamefont {Hur}}, \bibinfo {author} {\bibfnamefont
  {M.}~\bibnamefont {Arita}}, \bibinfo {author} {\bibfnamefont
  {K.}~\bibnamefont {Shimada}}, \bibinfo {author} {\bibfnamefont
  {H.}~\bibnamefont {Namatame}}, \ and\ \bibinfo {author} {\bibfnamefont
  {M.}~\bibnamefont {Taniguchi}},\ }\href {\doibase
  10.1103/PhysRevLett.108.046805} {\bibfield  {journal} {\bibinfo  {journal}
  {Phys. Rev. Lett.}\ }\textbf {\bibinfo {volume} {108}},\ \bibinfo {pages}
  {046805} (\bibinfo {year} {2012})}\BibitemShut {NoStop}%
\bibitem [{\citenamefont {Dresselhaus}(1955)}]{Dresselhaus1955}%
  \BibitemOpen
  \bibfield  {author} {\bibinfo {author} {\bibfnamefont {G.}~\bibnamefont
  {Dresselhaus}},\ }\href {\doibase 10.1103/PhysRev.100.580} {\bibfield
  {journal} {\bibinfo  {journal} {Phys. Rev.}\ }\textbf {\bibinfo {volume}
  {100}},\ \bibinfo {pages} {580} (\bibinfo {year} {1955})}\BibitemShut
  {NoStop}%
\end{thebibliography}%
\end{document}